\documentclass[twocolumn, times]{aastex62}
\usepackage{CJK}
\usepackage{graphicx}	
\usepackage{amsmath}	
\usepackage{amssymb}	
\usepackage{bm}		
\usepackage{txfonts}

\graphicspath{{./}{figures/}}

\received{?}
\revised{?}
\accepted{December 3, 2018}
\submitjournal{PASP}

\shorttitle{Disk instabilities}
\shortauthors{Sui et al.}

\begin{document}
\title{On the Gravitational Instabilities of Protoplanetary Disks}

\correspondingauthor{Min Li}
\email{min.li@unlv.edu, minphyli@163.com}

\author[0000-0003-0968-1554]{Ning Sui}
\affil{College of Physics, Jilin University, Changchun, Jilin 130012, China}
\affil{Max-Planck-Institut f\"{u}r Astronomie, K\"{o}nigstuhl 17, 69117  Heidelberg, Germany}

\author{Ping He}
\affil{College of Physics, Jilin University, Changchun, Jilin 130012, China}
\affil{NAOC–UKZN Computational Astrophysics Centre (NUCAC), University of KwaZulu-Natal, Durban 4000, South Africa}
\affil{Center for High Energy Physics, Peking University, Beijing 100871, China}

\author[0000-0002-9736-6978]{Min Li}
\affil{Department of Physics and Astronomy, University of Nevada, Las Vegas, NV 89154, USA}



\begin{abstract}

The gravitational instabilities are important to the evolution of the disks and the planet formation in the disks. We calculate the evolution of the disks which form from the collapse of the molecular cloud cores. By changing the properties of the cloud cores and the hydrodynamical viscosity parameters, we explore their effects on the properties of the gravitational instabilities. We find that the disk is unstable when the angular velocity of the molecular cloud core is larger than a critical value. The time duration of the instability increases as the angular velocity of the core increases. The increase of the hydrodynamical viscosity parameter hardly affects the stability of the disk, but decreases the time duration of the critical state of the gravitational instability in the disk. The instability of the disks can happen at very early time of evolution of the disk, which is consistent with the observations.

\end{abstract}


\section{Introduction}

The gravitational instabilities of the protoplanetary disks are
significant in the evolution of the disks.
The instability contributes to the transport of angular momentum in the disk
\citep{tom94,bod95,rud95,dur07,for11,arm11,ric11}.
FU Orionis events which occur at protostar stage \citep{bel95}
can be explained by the gravitational instability \citep{arm01}.
Gravitational instability also plays an important role in the formation of planetesimals, gas giant planets, brown dwarf companions,
and binary stars \citep{lau96,may04,mat05,ric06,bos11,bos12}.
Observations show that there are massive disks around very young protostars \citep{oso03,rod05,eis05},
which suggests that the disks can experience gravitational instability at early stage of evolution.

It has been suggested that the properties of protoplanetary disks
are related to the properties (masses, temperatures, and angular velocities) of the initial molecular cloud cores
\citep{vor09,vorb09,vor10a,vor10b,vorb10,vor11,vor13,jin14,li15,li16,li17}.
There is a link between the stability of the disk and the properties of the cloud core.
By using specific initial surface densities of disks to stand for the molecular cloud cores,
Vorobyov and Basu investigated the relation of the properties of cores and the disks in a series of papers.
\citet{vor10b} showed that a cloud core with high mass or high angular momentum leads to a disk which is more likely to be unstable.
\citet{vorb10} gave the same conclusion and further showed that
high temperature of the cloud core moderates the fragmentation of the resulting disk.

The instability of the disks also relates to the viscosities in the disks
which drive the evolution of the protoplanetary disks \citep{bal91,sui16}.
Generally, the viscosities are parameterized by the $\alpha$-prescription \citep{sha73}.
Usually, the $\alpha$-parameters are considered to be constant with time and radius.
Using these kinds of disks, \citet{vorb09} showed that if the viscosity parameter $\alpha$ increases,
the disk becomes more gravitationally stable.

In \citet{jin10}, the $\alpha$-parameters change with radius and time.
In the middle region of the disk, the ionization is weak and
the magnetorotational instability (MRI) \citep{bal91} does not work (dead zone).
The disk evolution in this region is driven by the hydrodynamic processes
\citep{dub93,fle03,kla03,dub05,cha06}.
The exact value of the hydrodynamical viscosity parameter is not well determined.
The change of the hydrodynamical viscosity will change the evolution of the disk.
Therefore, it is important to show the effects of the hydrodynamical viscosity
on the stabilities of the disks.

In this paper, we adopt a disk evolution model in which the disk forms from the collapse of
a molecular cloud core \citep{jin14,li15}.
We change the angular velocity of the cloud core more continuously
and show gravitational stability of the disk and the change of the time duration of the instability.
In our disk, the $\alpha$-parameter changes with radius and time and there is
a dead zone in the middle region of the disk \citep{jin10,jin14,li15}.
We also change the hydrodynamic viscosities and give the impacts of the
changes on the stabilities of the disks.


\section{MODEL DESCRIPTION} \label{model}

\subsection{disk evolution}
We consider axially symmetric and thin disks. In this case, the evolution of a disk is \citep{jin14,li15}
\begin{equation}\label{equ.diff}
\begin{aligned}
\frac{\partial \Sigma(R,t)}{\partial t}
=&\frac{3}{R} \frac{\partial}{\partial R} \left[ R^{1/2} \frac{\partial}{\partial R} (\Sigma \nu R^{1/2}) \right] +S(R,t)\\
&+S(R,t)\left\{2-3\left[\frac{R}{R_{d}(t)}\right]^{1/2}+\frac{R/R_{d}(t)}{1+[R/R_{d}(t)]^{1/2}}\right\},
\end{aligned}
\end{equation}
where $\Sigma(R,t)$ is the gas surface density of the disk at radius $R$ and time $t$,
$\nu$ is the kinematic viscosity,
$S(R,t)$ is the mass influx from the collapse of the cloud core onto the disk and protostar system,
and $R_{d}(t)$ is the centrifugal radius \citep{nak94}.
{The third term on the right-hand side is due to the difference of the specific angular momentum between the infall material and that in the disk.}
Solving the evolution equation (1) from the onset of the collapse,
we can get the surface density of the disk at any radius and at any time.

The turbulence viscosity can be written in the form
\begin{equation} \label{equ.nu}
\nu=\alpha c_s h,
\end{equation}
where $\alpha$ is the dimensionless Shakura-Sunyaev
parameter \citep{sha73},
$c_s$ is the sound speed in the mid-plane of the disk, which is
\begin{equation} \label{equ.cs}
c_s=\left(\Re T_m/\mu\right)^{1/2},
\end{equation}
where $\Re$ is gas constant, $T_m$ is temperature of the mid-plane of the disk,
and $\mu=2.33$ is the mean molecular weight,
and $h$ is the half thickness of the gas disk.

When the disks are gravitationally unstable, the MRI is believed to be the mechanism for transfer of angular momentum in the disks if there is a weak magnetic field \citep{bal91}. For ideal MHD, an effective $\alpha$ value is around $10^{-2}$ \citep[see e.g.,][]{dav10}. But simulations of non-ideal MHD for disks show that both Ohmic resistivity and ambipolar diffusion suppress the MRI in certain regions of disks. Ohmic resistivity is important around the midplane \citep{fle03}, while ambipolar diffusion is effective in the atmospheres of the disks \citep{bai13}. Moreover, \citet{les14} show that magnetic fields are generated in the midplane when considering the Hall effect. In summary, both ambipolar diffusion and Hall effect change the conventional layered accretion model of the disks. Global simulations are being done to find a better understanding of the disks \citep{gre15,bet16,bai17}. Given this complexity, we use a conventional layered model and the numerical results of the simulation for Ohmic resistivity \citep{fle03}. We also calculate the thermal ionisation rate in the inner regions of the disks. \citet{bai13} show that an conventional layered accretion model without ambipolar diffusion could give reasonable disk accretion rates.

If the disks are gravitationally stable,
the disks can be divided into three regions
according to the different features of the viscosity.
In the inner region,
the MRI may survive due to the thermal ionization.
We calculate the ionization from the results of \citet{ume83}.
{The ionization is a function of the surface density and the temperature in the disk.
We get the ionization fraction $x=n_e/n_{\rm H}$ from fitting the curve in Figure 7 of \citet{ume83},
where $n_e$ is the number density of electrons and $n_{\rm H}$ is the number density of hydrogen.
Thus the resistivity can be calculated from}
\begin{equation} \label{equ.eta}
\eta=6.5\times10^3x^{-1}\ \rm{cm^2\ s^{-1}}.
\end{equation}
{Then the magnetic Reynolds number is}
\begin{equation} \label{equ.Rem}
{\rm{Re}}_{\rm m}=\frac{c_s^2}{\eta\Omega}.
\end{equation}
{Here $\Omega$ is Keplerian angular velocity which is given by
$\Omega=\sqrt{G M_*/R^3}$, where $G$ is the gravitational
constant and $M_*$ is the central star mass.
The relation of $\alpha$ and ${\rm{Re}}_{\rm m}$ can be get from Table 1 of \citet{fle03}}.
In the outer region of the disk,
{the surface densities are low so that the cosmic rays can penetrate.}
We use $\alpha=0.008$ to represent the MRI caused by the cosmic rays.
In the intermediate region of the disk,
the MRI is active only at the surface of the disk due to the cosmic penetration.
The depth of the penetration and the $\alpha$ value can be calculated from \citet{fle03}.
In the middle plane of the intermediate region,
the MRI does not survive {(Dead zone)}.
The disk evolves under the effects of the hydrodynamical viscosities.
From the numerical simulations, the value of the hydrodynamic viscosity, $\alpha_{\rm hy}$,
is in the range from $10^{-5}$ to $0.1$ \citep{cha06,kla03,nak94},
while the exact value has not been determined yet \citep{dub93,fle03,kla03,dub05,cha06}.
We take $\alpha_{\rm hy}$ as a parameter.
{The $\alpha$ values are functions of the surface density and the temperature in the disks.
As the disk evolves with time, the boundaries of the regions and the $\alpha$ values change.}

{In general, $\alpha$ value can be summarized as follows:}
\begin{equation}
\alpha=\left\{
             \begin{array}{ll}
             0.02, &  \text{{gravitationally unstable;}}\\
             \alpha\left(\rm{Re}_{\rm m}\right), &\text{{gravitationally stable, }} \\
             &\text{{thermal ionization, inner region;}}\\
             \alpha\left(\rm{Re}_{\rm m}\right), &\text{{gravitationally stable, }} \\
             &\text{{cosmos rays penetrate the}}\\
             &\text{{surface of the intermediate region;}}\\
             \alpha_{\rm hy},&\text{{gravitationally stable, }} \\
             &\text{{dead zone, intermediate region;}}\\
             0.008, & \text{{gravitationally stable, cosmos rays }}\\
              & \text{{penetrate the outer disk,}}\\
             \end{array}
\right.
\end{equation}
{where $\alpha(\rm{Re}_{\rm m})$ is obtained from the results of the non-ideal MHD simulations given by \citet{fle03}.}

The temperature of the disk is calculated by assuming local balance of heating and cooling
at the surface of the disk. The temperature at the surface and the midplane of the disk are \citep{nak94,hue05}
\begin{equation}
\sigma T_{s}^{4}=\frac{1}{2}
\left(1+\frac{1}{2\tau_{P}}\right)(\dot{E_{\nu}}+\dot{E_{s}})+\sigma
T_{\rm ir}^{4} +\sigma
T_{\rm acc}^{4} +\sigma T^{4},
\end{equation}
and
\begin{equation}
\begin{aligned}
\sigma T_{m}^{4}=&\frac{1}{2}
\left[\left(\frac{3}{8}\tau_{R}+\frac{1}{2\tau_{P}}\right)\dot{E_{\nu}}
+\left(1+\frac{1}{2\tau_{P}}\right)\dot{E_{s}}\right] \\
&+\sigma T_{\rm ir}^{4}+\sigma T_{\rm acc}^{4} +\sigma T^{4}.
\end{aligned}
\end{equation}
Here $\sigma$ is the Stefan-Boltzmann constant,
$\tau_{P}=\kappa_{P}\Sigma$ is the Planck mean optical depth,
where $\kappa_{P}$ is the Planck mean opacity,
$\dot{E_{\nu}}$ is the viscous dissipation rate,
$\dot{E_{s}}$ is the energy generation rate by shock heating,
$T_{\rm ir}$ is the effective temperature due to
the irradiation from the protostar,
{$T_{\rm acc}$ comes from the luminosity due to the mass accretion from the disk onto the central star \citep{cas94},}
{$T$ is the temperature of the molecular cloud core,}
$\tau_{R}=\kappa_{R}\Sigma$ is the Rosseland mean optical
depth, and $\kappa_{R}$ is the Rosseland mean opacity.
Approximately,
$\kappa_{P}=2.39\kappa_{R}$ \citep{nak94}.

The disk becomes gravitationally unstable, i.e., when the Toomre parameter
$Q$ \citep{too64}, which is given by
\begin{equation} \label{equ.q}
Q=\frac{c_s \Omega}{\pi G \Sigma},
\end{equation}
is less than $Q_{\rm crit}$ (of the order of unity) and we adopt
$Q_{\rm crit}=1$. Therefore,
decreasing the temperature of the disk, which results in
the decrease of $c_s$ (see Equation (\ref{equ.cs})), and increasing $\Sigma$ and $R$ favor
the instability of the disk.
If $Q$ is less than 1, we use the results of \citet{lau94}, \citet{lau96}, \citet{lau97}, and \citet{lau98}
and adopt $\alpha=0.02$ in the disk.

\subsection{Initial conditions}
Observationally, a molecular cloud core can be characterized by temperature ($T$),
mass ($M$), and angular velocity ($\omega$). \citet{jij99} reviewed the temperatures
of the cloud cores and showed that the median value is $\sim 15\ \rm K$.
We take it as our typical value. The mass we adopt is $1\ \rm M_{\odot}$.
\citet{goo93} analyzed the angular velocities of the cloud cores.
They found that the cores rotate rigidly.
The angular velocity ranges from $0.3$ to $13 \times 10^{-14}\ \rm s^{-1}$ \citep{shu77,goo93,jin10}.
{ The ratio of rotational-to-gravitational energy of the pre-stellar cores is
\begin{equation}
    \beta=\frac{1}{16}\frac{p}{q}\frac{G^2 \mu^{3}}{\Re^{3}}\frac{M^2 \omega^2}{T^{3}},
	\label{eq:beta2}
\end{equation}
where $p/q=0.22$ for a rigid rotating core with $\rho\propto r^{-2}$.
Hence the corresponding ratios of rotational to gravitational energy $\beta$ range from
 $4.31\times 10^{-5}$ to $8.09 \times 10^{-2}$}.
We adopt the values of angular velocities at fixed intervals.
When they are less than $1 \times 10^{-14}\ \rm s^{-1}$,
the interval is $0.1 \times 10^{-14}\ \rm s^{-1}$,
while it is $1 \times 10^{-14}\ \rm s^{-1}$ when they are larger than
$1 \times 10^{-14}\ \rm s^{-1}$.
For $\alpha_{\rm hy}$, we adopt $\alpha_{\rm hy}=5\times 10^{-4}$
and $\alpha_{\rm hy}=5\times 10^{-3}$.

{The inner boundary of our disk is 0.3 AU. The outer boundary is $1.25\times10^5$ AU,
which allows the disk to expand freely.
We use 281 logarithmically cells in the radial direction.
Both the boundaries and the cells in the disk are fixed at their initial values.}

\section{Effects of $\omega$}

We first investigate the effects of the angular velocities of the cloud cores,
$\omega$, on the evolution of the disks.
We fix $\alpha_{\rm hy}$ to be $5\times 10^{-4}$.
The mass and the temperature of the cloud cores
are $M=1.0\ M_{\odot}$ and $T=15\ \rm K$ respectively.

\subsection{Evolution of the surface density}

   \begin{figure}
   \centering
   \includegraphics[width=8cm]{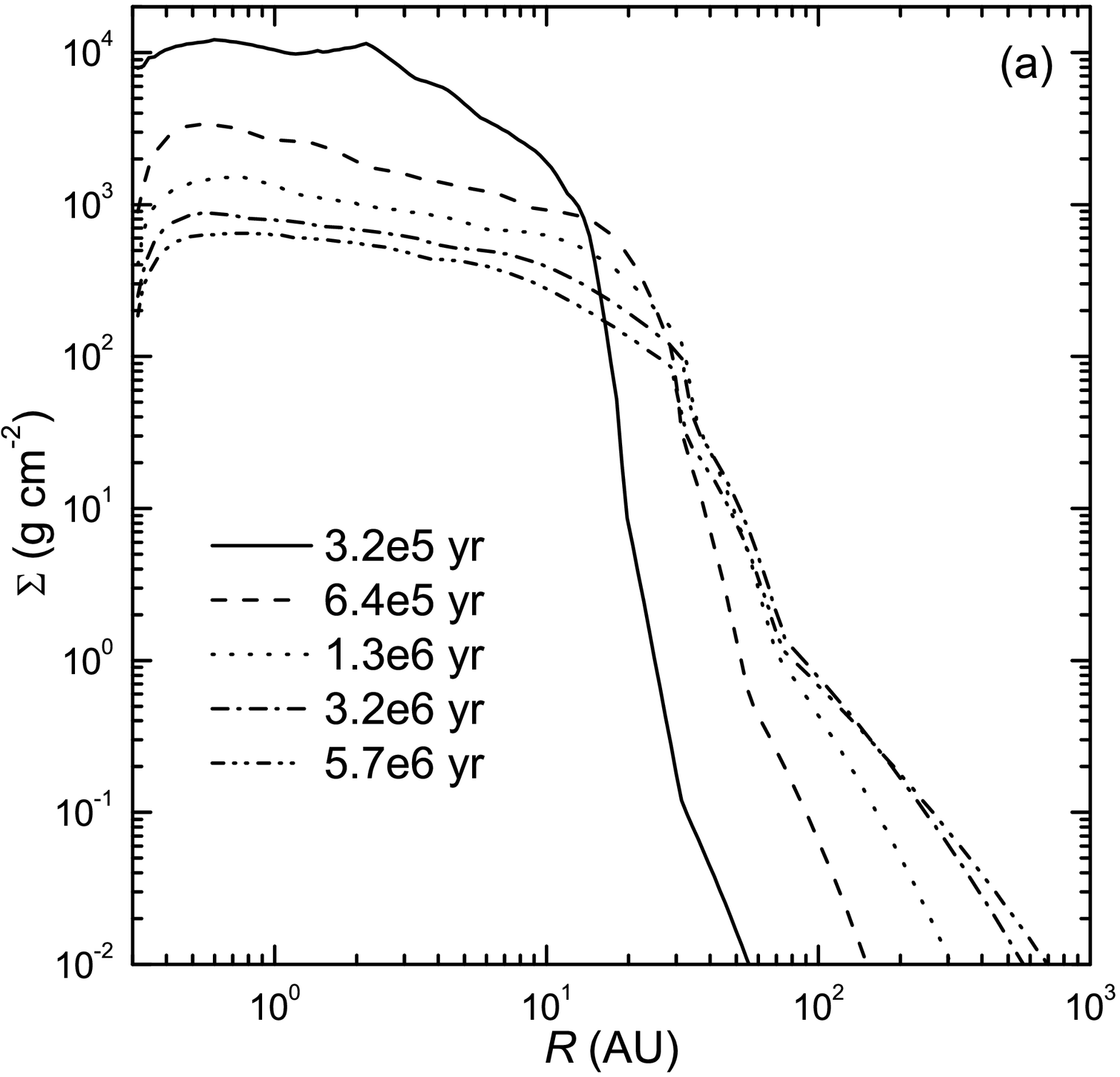}
   \includegraphics[width=8cm]{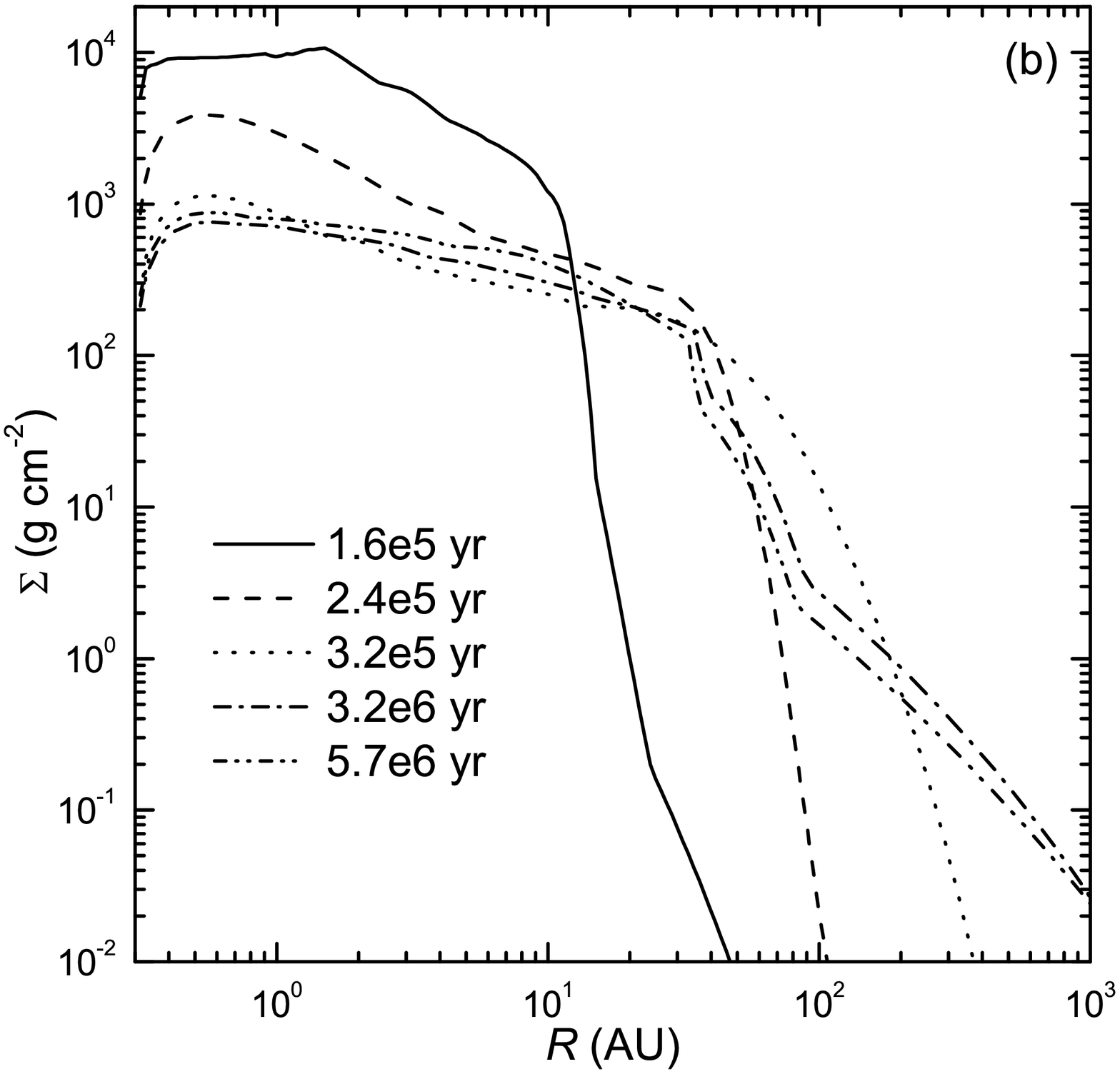}
      \caption{
      The evolution of surface density of the gas disk for two cases:
      (a) $\omega=1 \times 10^{-14}\ \rm s^{-1}$ and
      (b) $\omega=3 \times 10^{-14}\ \rm s^{-1}$.
              }
         \label{fig-sur}
   \end{figure}

We choose $\omega$ to be
$1 \times 10^{-14}\ \rm s^{-1}$ and $3 \times 10^{-14}\ \rm s^{-1}$.
The evolution of the surface density of the gas disk is shown in Fig. \ref{fig-sur}.
Fig. \ref{fig-sur} (a) shows the case for $\omega=1 \times 10^{-14}\ \rm s^{-1}$.
At the early evolution of the disk, the materials concentrate at the inner region {(within $\sim$ 10 AU)} of the disk.
As the disk evolves due to the viscosities,
the disk expands to large radius and
the surface density in the inner region decreases while that in the outer region increases.
Fig. \ref{fig-sur} (b) shows the case for $\omega=3 \times 10^{-14}\ \rm s^{-1}$.
The general trends of the evolution are similar to that in Fig. \ref{fig-sur} (a).
The difference is as follows.
As the angular velocity increases,
the angular momentum of the system increases,
the surface density in the inner region decreases more quickly and
there are more materials in the outer region.
For example, it takes $\sim10^6$ years for the surface densities within $\sim$ 10 AU
changing from $\sim 10^4$ to
$\sim 10^3 \rm \ g \ cm^{-2}$
for $\omega=1 \times 10^{-14}\ \rm s^{-1}$,
while it only takes about $1.6\times10^5$ years for $\omega=3 \times 10^{-14}\ \rm s^{-1}$
(See the first and the third lines in Fig. \ref{fig-sur} (a) and (b)).
The surface density at inner radii at $t=5.7\times10^6$ years is larger than that at $t=3.2\times10^6$
because of the inward movement of materials at radius larger than 20 AU.

\subsection{Evolution of the temperature}

\begin{figure}
   \centering
   \includegraphics[width=8cm]{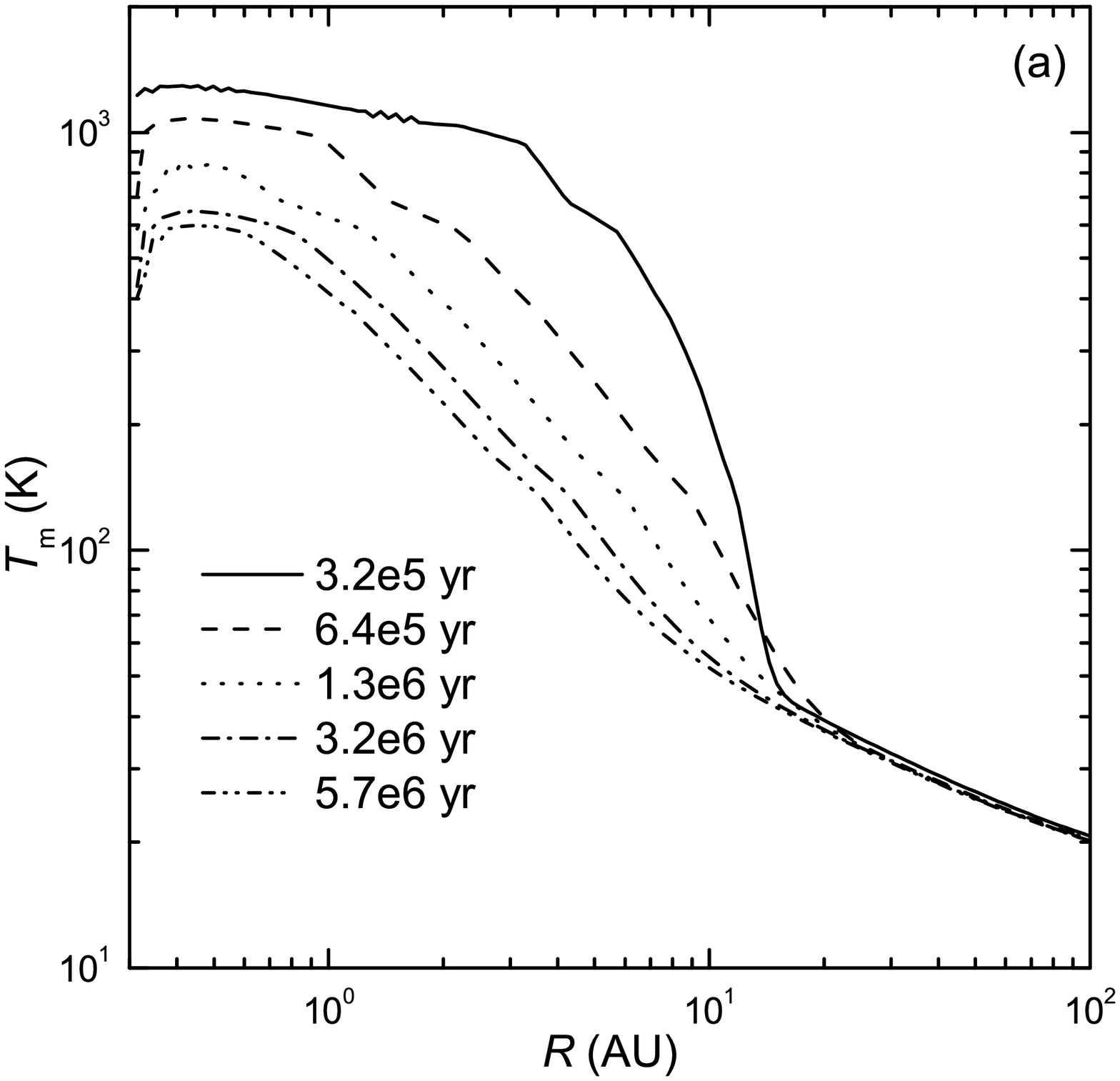}
   \includegraphics[width=8cm]{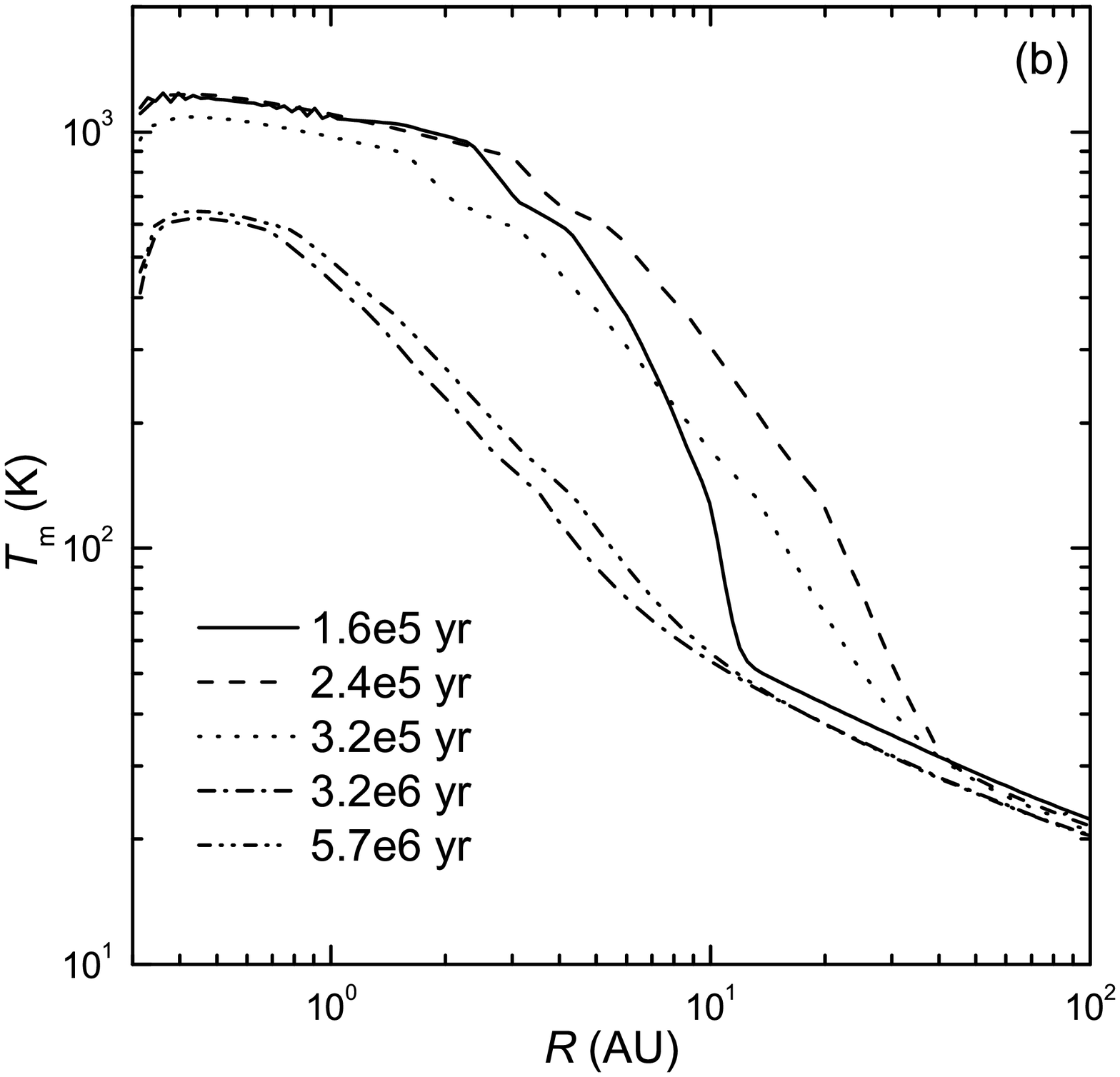}
      \caption{
      The evolution of the midplane temperature of the disks for two cases:
      (a) $\omega=1 \times 10^{-14}\ \rm s^{-1}$ and
      (b) $\omega=3 \times 10^{-14}\ \rm s^{-1}$.
      }
         \label{fig-tem}
\end{figure}

Fig. \ref{fig-tem} shows the evolution of the temperature of the disks.
Fig. \ref{fig-tem} (a) shows the case for $\omega=1 \times 10^{-14}\ \rm s^{-1}$ and
Fig. \ref{fig-tem} (b) is the case for $\omega=3 \times 10^{-14}\ \rm s^{-1}$.
For $\omega=1 \times 10^{-14}\ \rm s^{-1}$, the temperature decreases with radius and time, and
the decrease of the temperature in the inner region
is quicker than that in the outer region.
For higher $\omega$,
{the temperature increases due to the gravitational instability and accretion luminosity first and then
decreases due to the decrease of the surface densities.
The temperature at $t=5.7\times10^6$ years is larger than that at $t=3.2\times10^6$ as
the surface densities (See Fig. \ref{fig-sur}).}

\subsection{{Evolution of $\alpha$}}

\begin{figure}
   \centering
   \includegraphics[width=8cm]{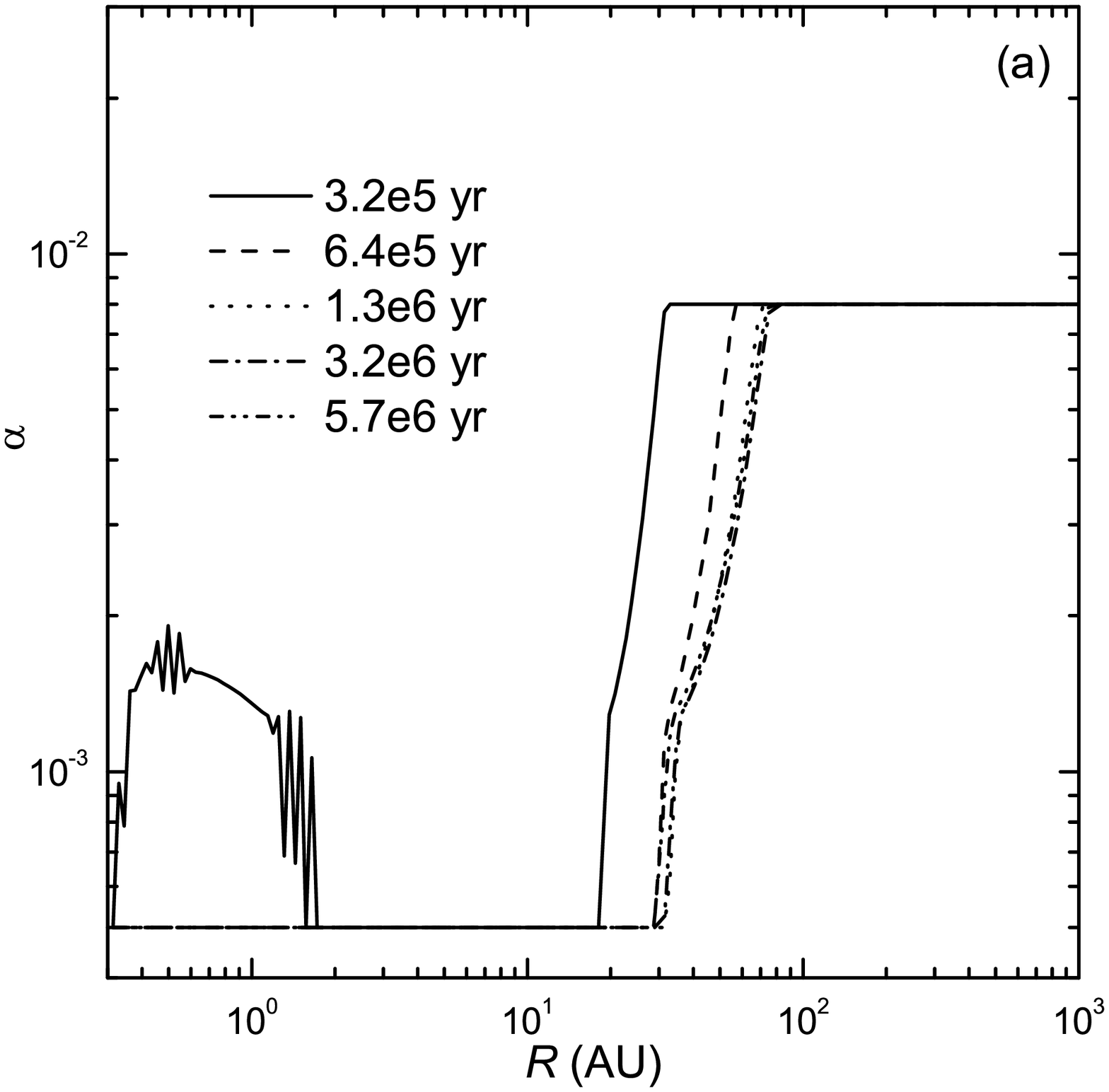}
   \includegraphics[width=8cm]{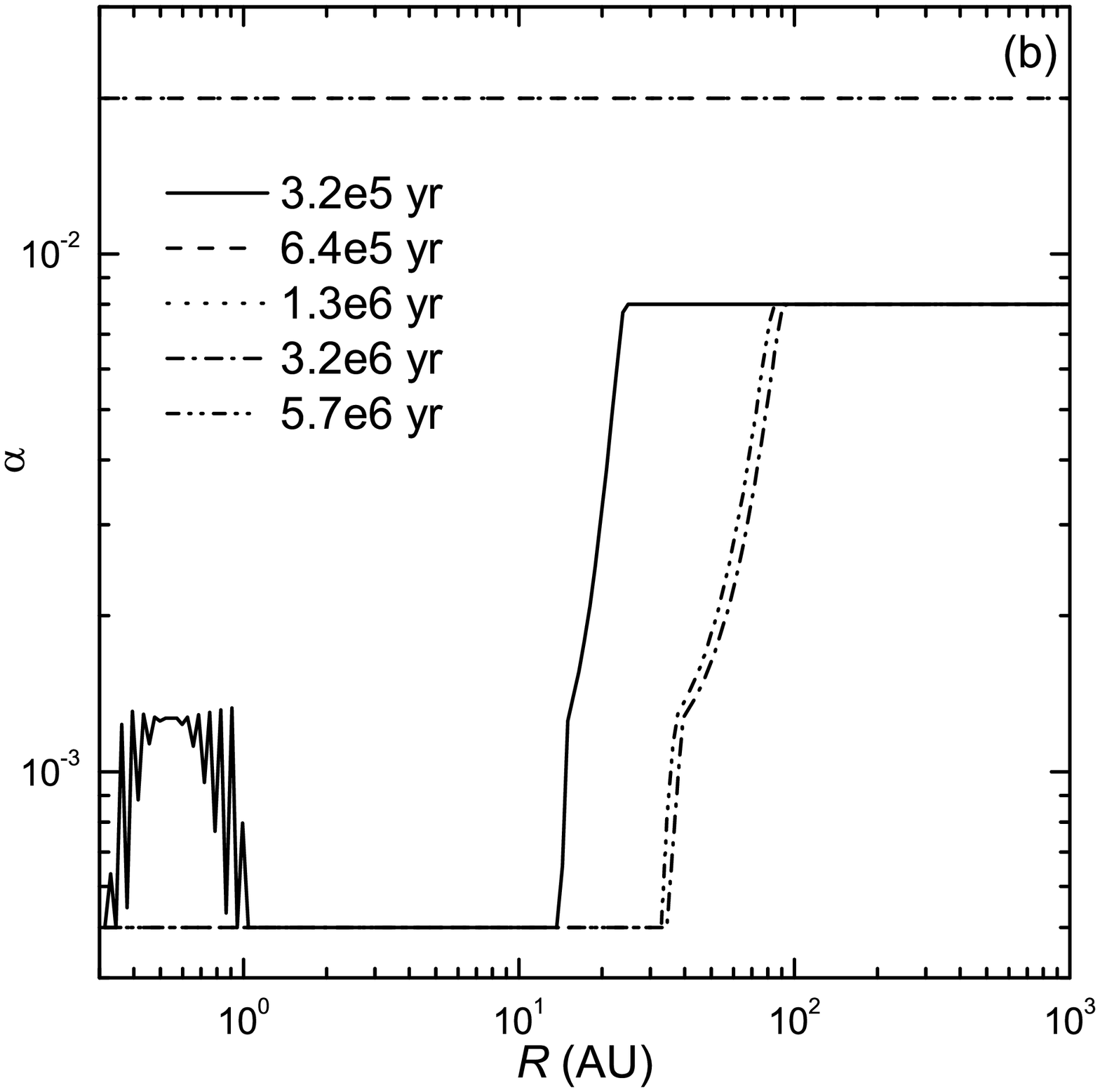}
      \caption{
{      The evolution of the $\alpha$ parameter in the disks for two cases:
      (a) $\omega=1 \times 10^{-14}\ \rm s^{-1}$ and
      (b) $\omega=3 \times 10^{-14}\ \rm s^{-1}$.
      In Case (b), when $t=6.4\times10^5$ and $t=1.3\times10^6$, the disk is unstable,
      $\alpha=0.02$.}
      }
         \label{fig-alpha}
\end{figure}

{Fig. \ref{fig-alpha} plots the evolution of $\alpha$ for
 $\omega=1 \times 10^{-14}\ \rm s^{-1}$ and
 $\omega=3 \times 10^{-14}\ \rm s^{-1}$.
 The thermal ionization only works at early evolution of the disk
 ($t=3.2\times10^5$) when the temperature is
 high. In the intermediate region, $\alpha$ is governed by the
 hydrodynamical viscosities.
 The outer radius of the intermediate region moves outward with time,
 which is caused by the expansion of the disk.
 In Case (b), the disk is unstable when $t=6.4\times10^5$ and $t=1.3\times10^6$,
$\alpha=0.02$.
 }

\subsection{Influence on the instability}
To show the dependence of the instability of the disks on $\omega$,
we change $\omega$ more continuously and run the corresponding evolution of the disks.
We find that a disk is stable when the angular velocity is less than
$1.2 \times 10^{-14}\ \rm s^{-1}$.

\begin{figure}
   \centering
   \includegraphics[width=8cm]{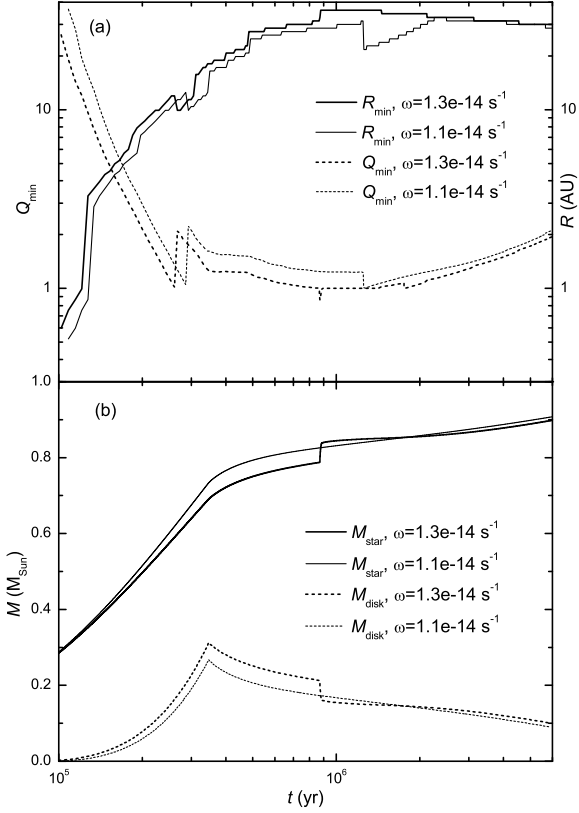}
      \caption{(a) The minimum value of $Q$ ($Q_{\rm min}$) and the corresponding radius ($R_{\rm min}$)
as functions of $t$ for two cases: $\omega=1.1 \times 10^{-14}\ \rm s^{-1}$
(lower solid line and upper short dashed line) and
$\omega=1.3 \times 10^{-14}\ \rm s^{-1}$
(upper solid line and lower short dashed line).
{(b) The corresponding evolution of the masses of the central stars and the disks. }
Here, $M=1\ M_{\odot}$, $T=15$ K, and $\alpha_{\rm hy}=5\times10^{-4}$.
              }
         \label{fig-q}
   \end{figure}

In Fig. \ref{fig-q} (a), we show $Q_{\rm min}$ (the minimum value of $Q$ of all the radius at a time) and
the corresponding radius ($R_{\rm min}$) as functions of $t$ for
$\omega=1.1 \times 10^{-14}\ \rm s^{-1}$
and $\omega=1.3 \times 10^{-14}\ \rm s^{-1}$.
Generally, $Q_{\rm min}$ decreases with time first due to the increase of surface density as the materials
fall onto the disk. And then $Q_{\rm min}$ increases quickly at $\sim3\times10^5$ year due to the quick decrease of surface density
after the infall ends. After that, $Q_{\rm min}$ decreases slowly
due to the combined effect of the decreasing surface density and temperature.
At last (after $\sim1\times10^6$ year), the temperature decreases slowly, and
$Q_{\rm min}$ increases slowly due to the decrease of surface density.

For the case of $\omega=1.1 \times 10^{-14}\ \rm s^{-1}$,
$Q_{\rm min}$ is greater than 1 at any time and at any radius.
So the disk is stable all the time.
The minimum value of $Q_{\rm min}$ is 1.002 at $t=1.26\times10^6$ yr
and the corresponding radius is 21.7 AU.
For the case of $\omega=1.3 \times 10^{-14}\ \rm s^{-1}$,
there are intervals that $Q_{\rm min}$ is less than 1.
The longest interval begins at $t=8.7\times 10^5$ yr and lasts about $1.3\times 10^4$ yr.
For most of the evolution, $Q_{\rm min}$ locates at around 30 AU.
Therefore, as $\omega$ increases, there are more materials at large radius
and the disk is inclined to trigger instability at large radius.

{Fig. \ref{fig-q} (b) shows the corresponding evolution of
the masses of the central stars and the disks.
For both cases, the masses of the central stars increase quickly due to the infall from
the molecular cloud cores and slowly after the infall ceases.
The masses of the disks also increase quickly during the infall stage and then decrease slowly
due to the accretion to the central star.
Before the gravitational instability onsets, the disk mass for the case of
$\omega=1.3 \times 10^{-14}\ \rm s^{-1}$ is higher than that of
$\omega=1.1 \times 10^{-14}\ \rm s^{-1}$.
The reason is that as the angular momentum of the system increases, there will be more material expanding to a large radius and less material accreted to the central star.
For the case of $\omega=1.3 \times 10^{-14}\ \rm s^{-1}$,
the disk becomes unstable when the disk mass is around 0.21 $M_\odot$.
When the disk is unstable, the disk mass decreases quickly while the stellar mass increases quickly.
The mass transferring speed for an unstable disk is higher than that for a stable disk.
Note that the disk does not become unstable when the disk mass reaches its maximum value.
This is because, at early evolution, the materials are mainly distributed at the small radius of the disk, the gravitational instability is not triggered.
}

\section{Effects of $\alpha_{\rm hy}$}

\subsection{Evolution of the surface density and the temperature}

\begin{figure}
   \centering
   \includegraphics[width=5.5cm]{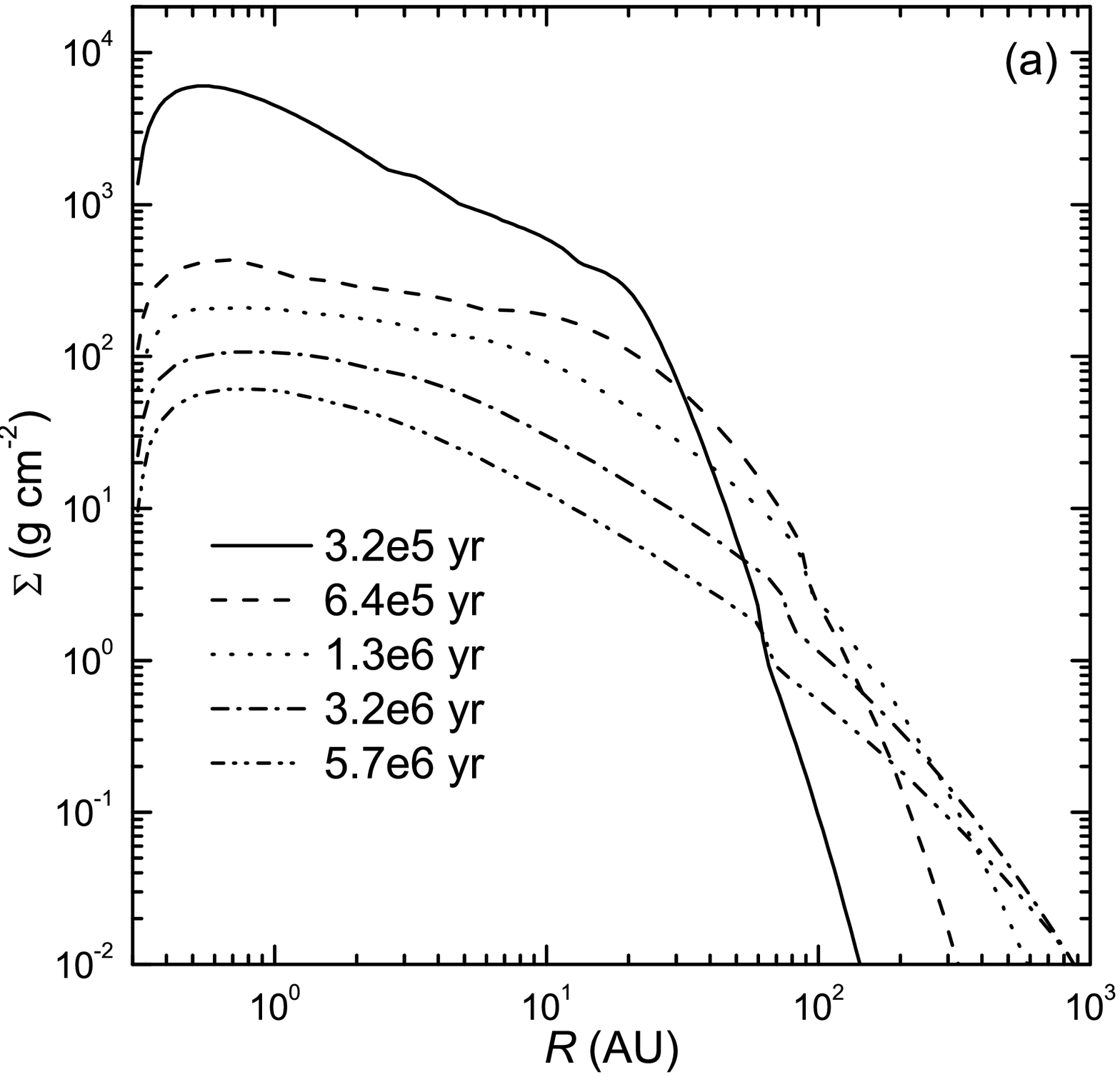}
   \includegraphics[width=5.5cm]{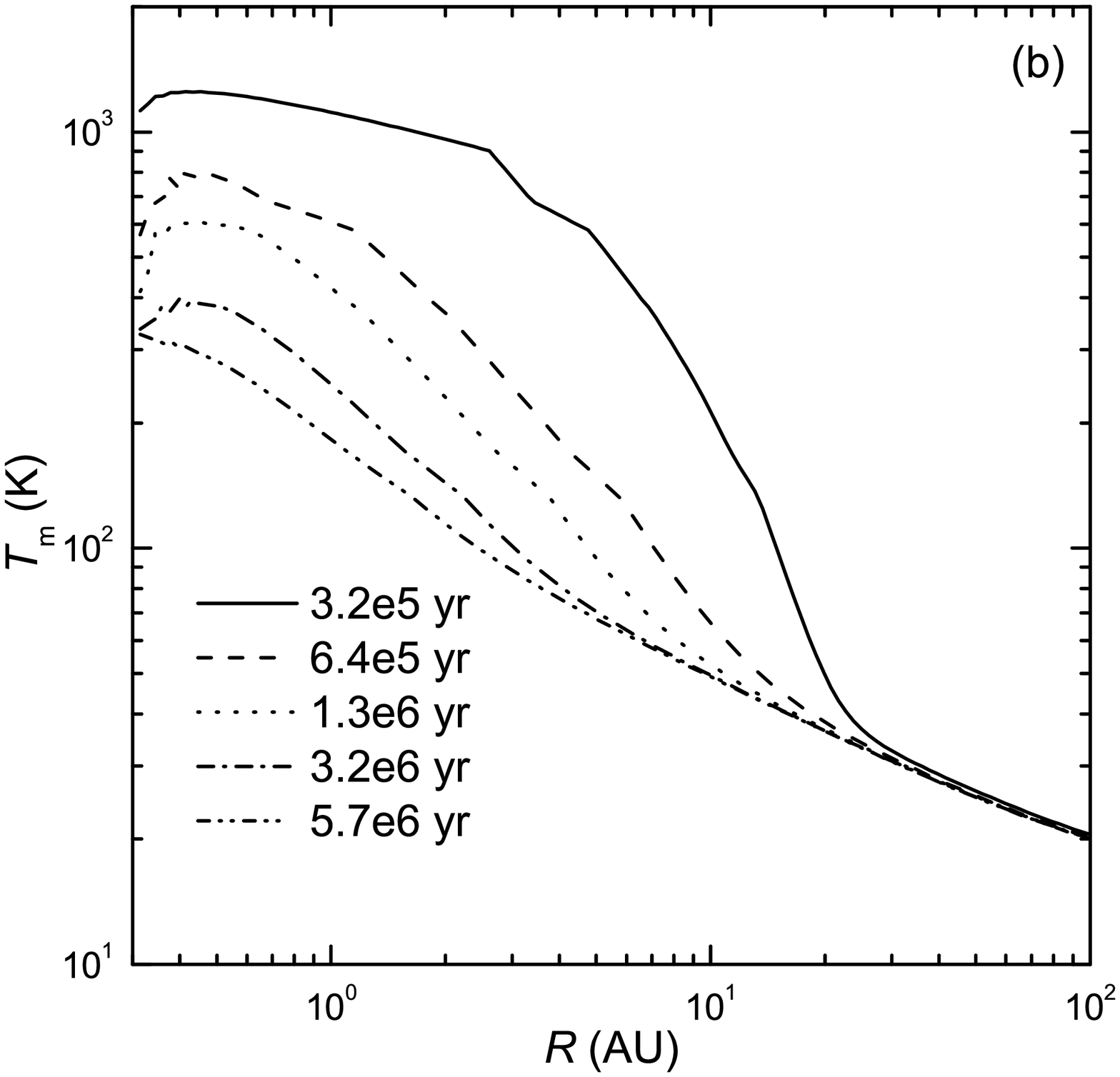}
   \includegraphics[width=5.5cm]{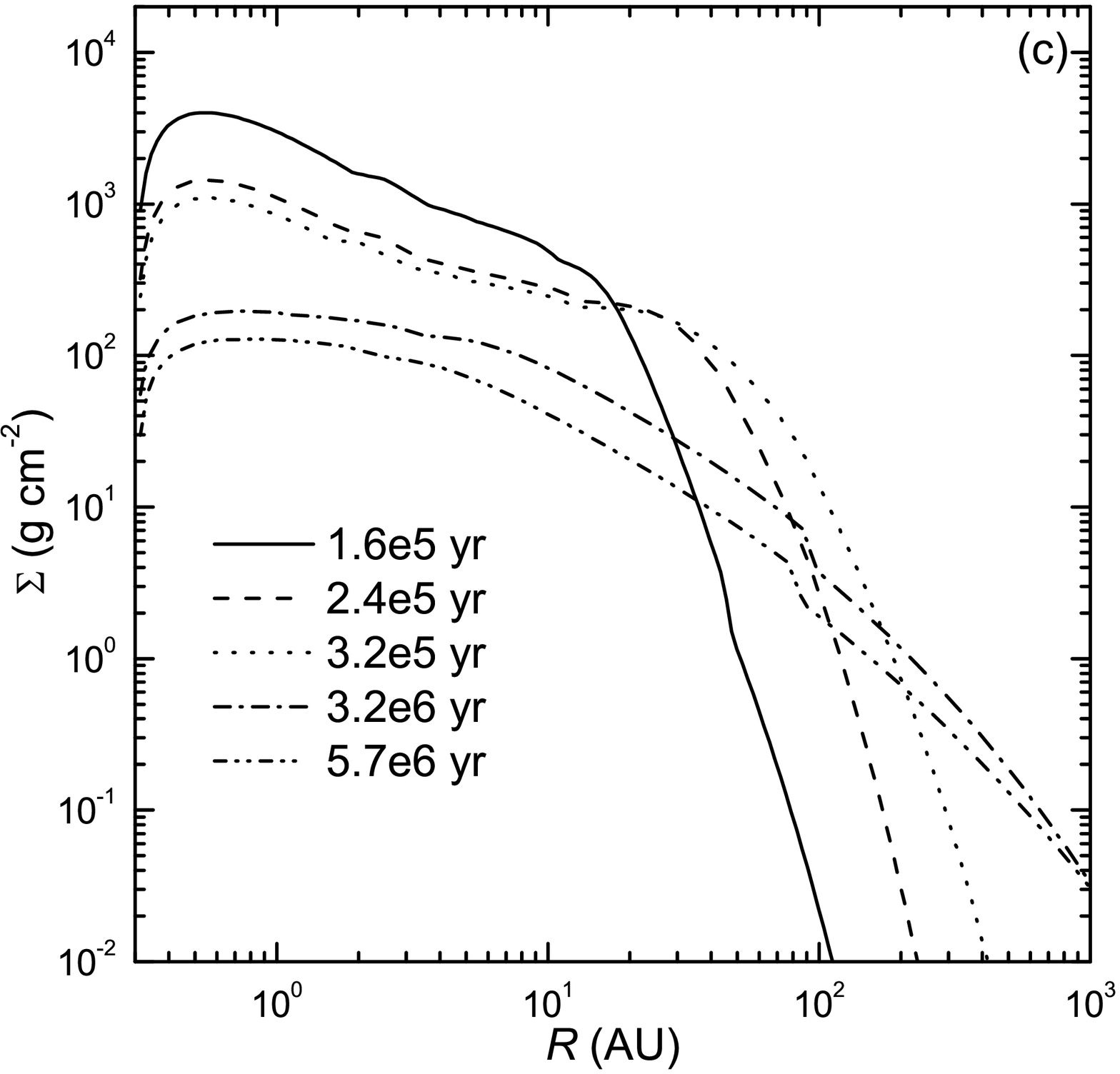}
   \includegraphics[width=5.5cm]{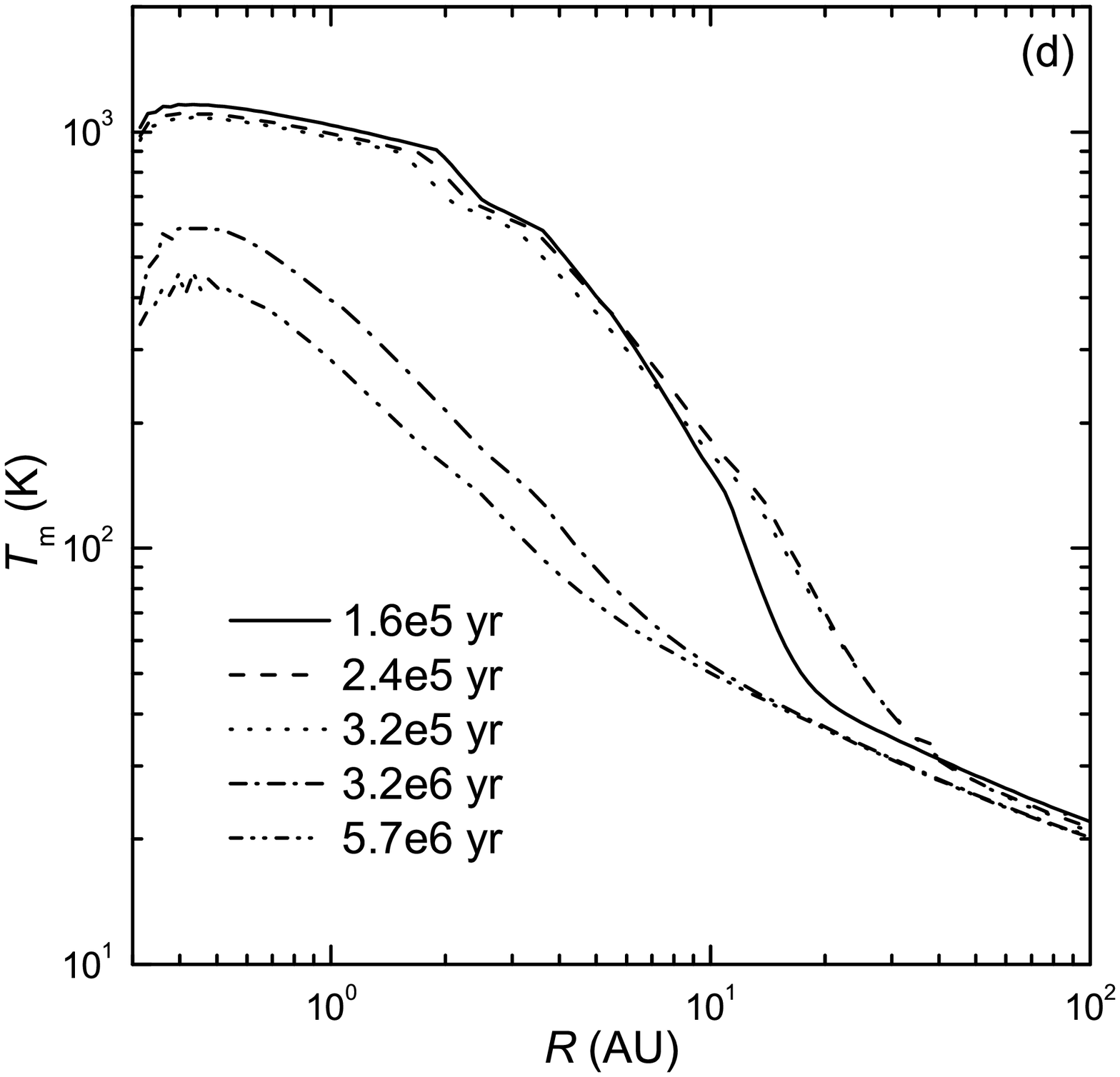}
      \caption{
      The evolution of surface density and the temperature of the gas disk for two cases with $\alpha_{\rm hy}=5\times 10^{-3}$.
      (a) and (b): $\omega=1 \times 10^{-14}\ \rm s^{-1}$, and
      (c) and (d): $\omega=3 \times 10^{-14}\ \rm s^{-1}$.
              }
         \label{fig-sur2}
   \end{figure}

We change $\alpha_{\rm hy}$ to be $5\times 10^{-3}$ and fixed
the temperature and the mass of the molecular cloud cores to be $T=15\ \rm K$ and $M=1\ M_{\odot}$.
For comparison, we plot the evolution of the surface density and the temperature with $\omega=1 \times 10^{-14}\ \rm s^{-1}$
and $\omega=3 \times 10^{-14}\ \rm s^{-1}$ for $\alpha_{\rm hy}=5\times10^{-4}$ and
$\alpha_{\rm hy}=5\times10^{-3}$ in Fig. \ref{fig-sur2}.
Generally, the surface density in the inner region decreases with time while in the outer region increases with time.
As the viscosity is larger, the surface density evolves more quickly compared to that in Fig. \ref{fig-sur} and the general density is lower
except that at the radius larger than $\sim$ 50 AU (Fig. \ref{fig-sur2} (a) and (c)).
The temperature of the disks decreases more quickly than that in the case
with $\alpha_{\rm hy}=5\times 10^{-4}$ (Fig. \ref{fig-sur2} (b) and (d)), which is caused mainly by the decrease of the surface density.

\subsection{Influence on the instability}
We also change $\omega$ continuously and run the corresponding evolution of the disks.
In this case, a disk begins to be unstable when
$\omega>1.2 \times 10^{-14}\ \rm s^{-1}$.
The critical value is almost the same as in
the case of $\alpha_{\rm hy}=5\times 10^{-4}$.
This can be explained as follows.
As the viscosity increases,
the gas in the disk will be accreted to the central star more quickly.
This leads to the decrease of the surface density of the disk
and favors the stability of the disk.
However, for the conservation of the angular momentum,
there will be more materials expanding to larger radius,
and this will favor the instability of the disk \citep{rud86}.
These two effects offset each other and
so the rise of the instability does not change much,
i.e., for all the angular velocities,
the stability of the disks does not change
when $\alpha_{\rm hy}$ changes from $5\times 10^{-4}$ to $5\times 10^{-3}$
for cloud cores with $T=15\ \rm K$ and $M=1\ M_{\odot}$.

\begin{figure}
   \centering
   \includegraphics[width=8cm]{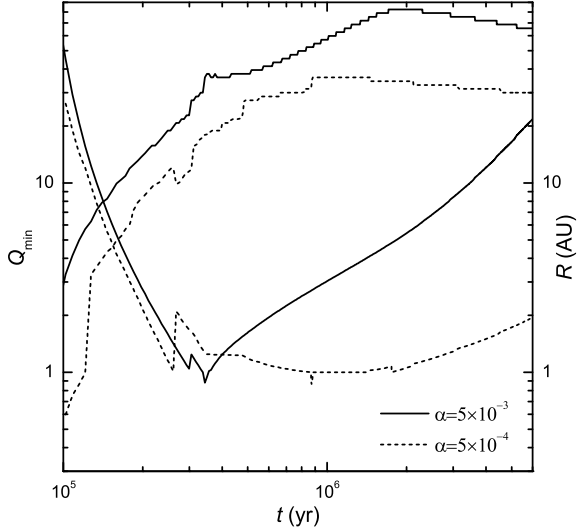}
      \caption{The minimum value of $Q$ ($Q_{\rm min}$) and the corresponding radius ($R_{\rm min}$)
as functions of $t$ for
$\omega=1.3 \times 10^{-14}\ \rm s^{-1}$
with $\alpha_{\rm hy}=5\times10^{-3}$.
The case for $\alpha_{\rm hy}=5\times10^{-4}$ and $\omega=1.3 \times 10^{-14}\ \rm s^{-1}$ is also plotted.
              }
         \label{fig-q50}
   \end{figure}

Fig. \ref{fig-q50} shows $Q_{\rm min}$ and $R_{\rm min}$ as functions of $t$ for
$\omega=1.3 \times 10^{-14}\ \rm s^{-1}$ and $\alpha_{\rm hy}=5\times 10^{-3}$.
For comparison, we also plot the case for $\omega=1.3 \times 10^{-14}\ \rm s^{-1}$ and $\alpha_{\rm hy}=5\times 10^{-4}$ in the same figure.

For $\alpha_{\rm hy}=5\times 10^{-3}$, first $Q_{\rm min}$ is larger than the case for $\alpha_{\rm hy}=5\times 10^{-4}$ when $t<\sim2.6\times10^5$ yr.
And then $Q_{\rm min}$ is smaller and there is interval that $Q_{\rm min}$ is less than 1 when $t\sim4.0\times10^5$ yr
(later than the case of $\alpha_{\rm hy}=5\times 10^{-4}$).
After that, $Q_{\rm min}$ increases quickly and is larger than 1 all the time.
For $\alpha_{\rm hy}=5\times 10^{-4}$, $Q_{\rm min}$ is around 1 when $t>\sim1\times10^6$ yr and $t<\sim2\times10^6$,
which means that the disk is in the critical state of the gravitational instability for this time duration.

Generally, $R_{\rm min}$ for $\alpha_{\rm hy}=5\times 10^{-3}$ is larger than the case for $\alpha_{\rm hy}=5\times 10^{-4}$
and is mostly larger than 50 AU, which is consistent with the radius at which the surface density for $\alpha_{\rm hy}=5\times 10^{-3}$ is larger than the case for $\alpha_{\rm hy}=5\times 10^{-4}$ (Fig. \ref{fig-sur2} vs. Fig. \ref{fig-sur}).
This is because, for $\alpha_{\rm hy}=5\times 10^{-3}$, there are more materials at larger radius,
and both the larger radius and the larger surface densities there favor the instability.

\subsection{Time durations of the instability}

   \begin{figure}
      \centering
   \includegraphics[width=8cm]{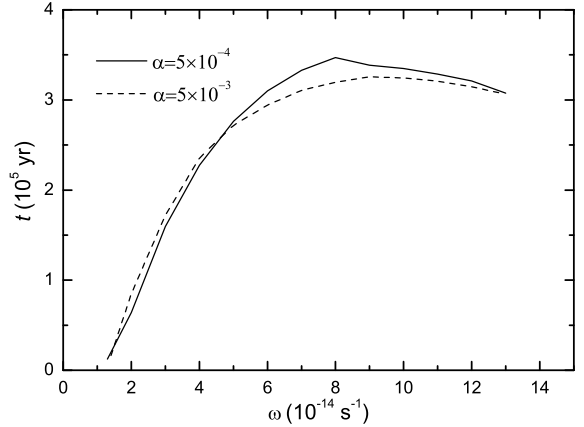}
      \caption{The time durations of the instabilities as functions of $\omega$ for $\alpha_{\rm hy}=5\times10^{-3}$
and $\alpha_{\rm hy}=5\times10^{-4}$.
              }
         \label{Figdur}
   \end{figure}

      \begin{figure}
   \centering
   \includegraphics[width=6cm]{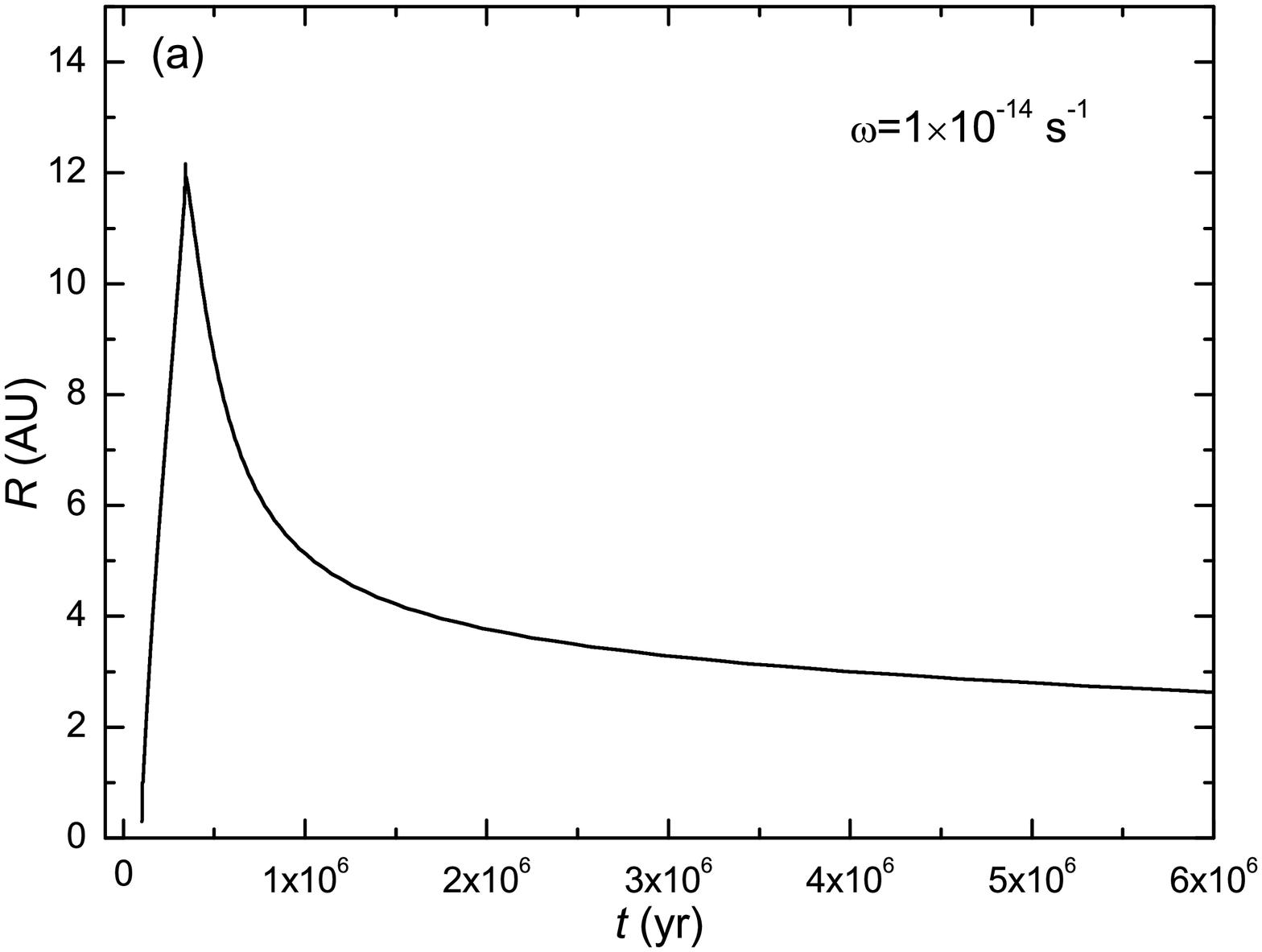}
   \includegraphics[width=6cm]{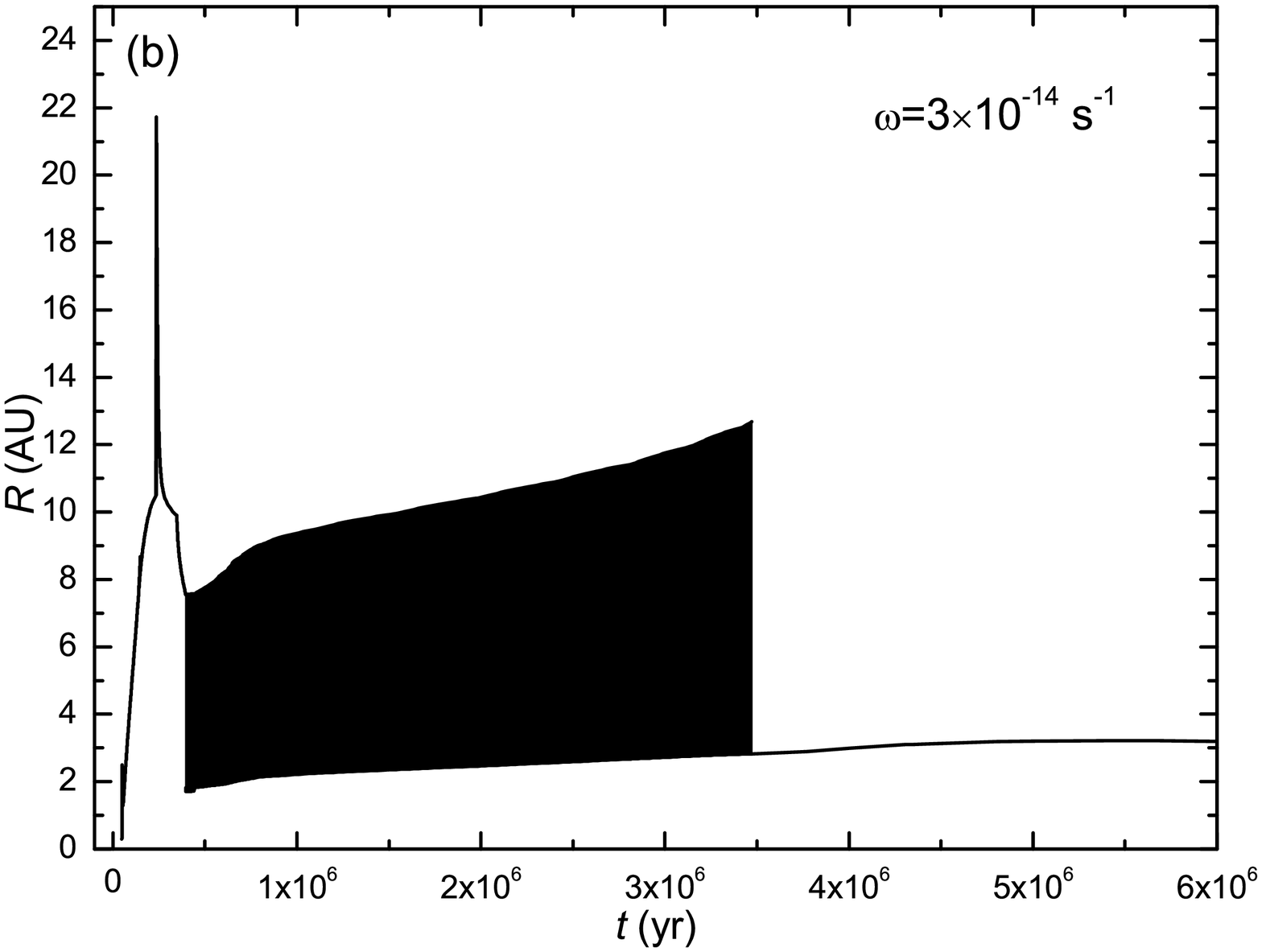}
   \includegraphics[width=6cm]{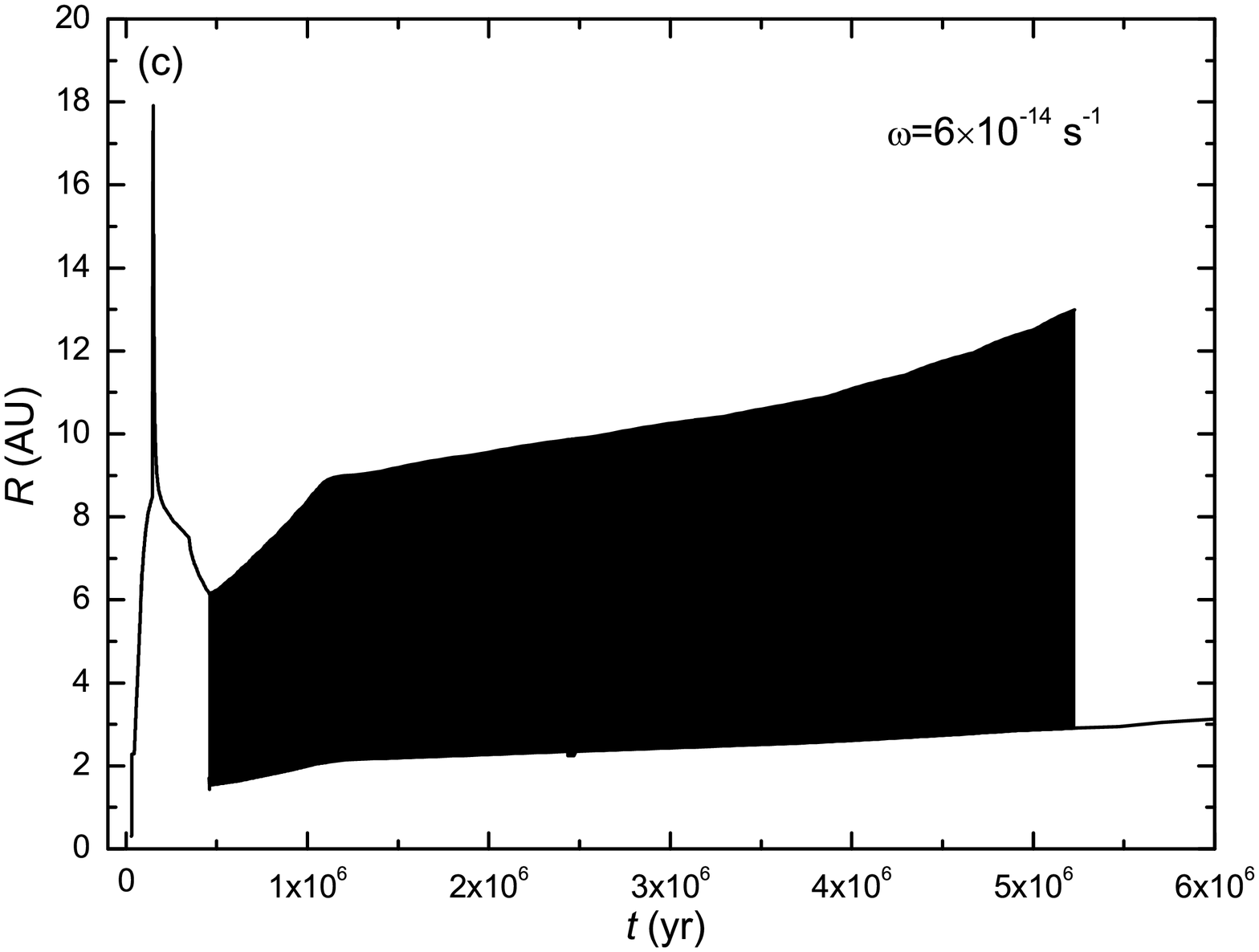}
   \includegraphics[width=6cm]{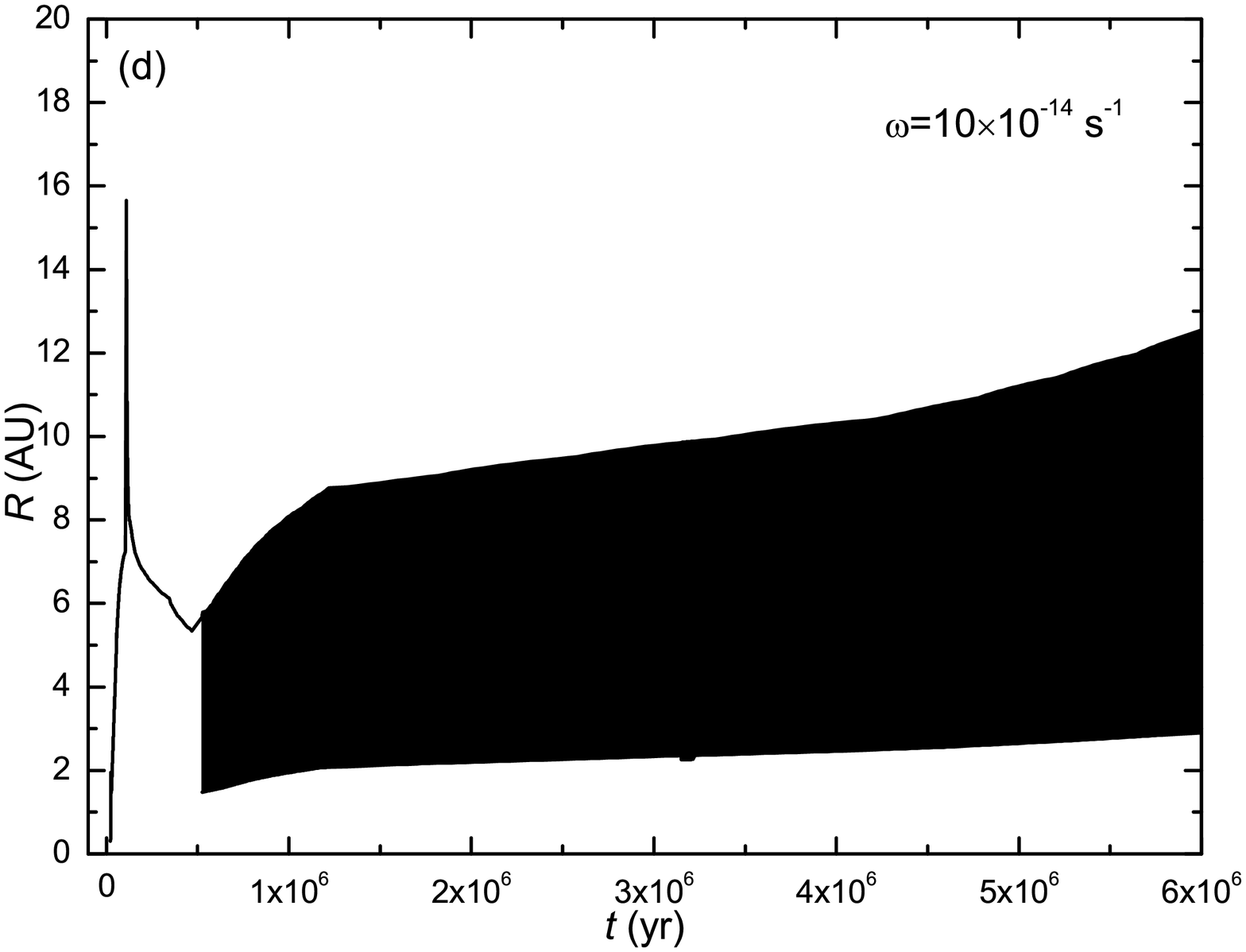}
      \caption{The evolution of the snowline for $\omega=1 \times 10^{-14}\ \rm s^{-1}$,
$\omega=3 \times 10^{-14}\ \rm s^{-1}$,
$\omega=6 \times 10^{-14}\ \rm s^{-1}$,
and $\omega=10 \times 10^{-14}\ \rm s^{-1}$, respectively.
Here $\alpha_{\rm hy}=5\times10^{-4}$.
               }
         \label{Figsnowline}
   \end{figure}

   \begin{figure}
   \centering
   \includegraphics[width=4cm]{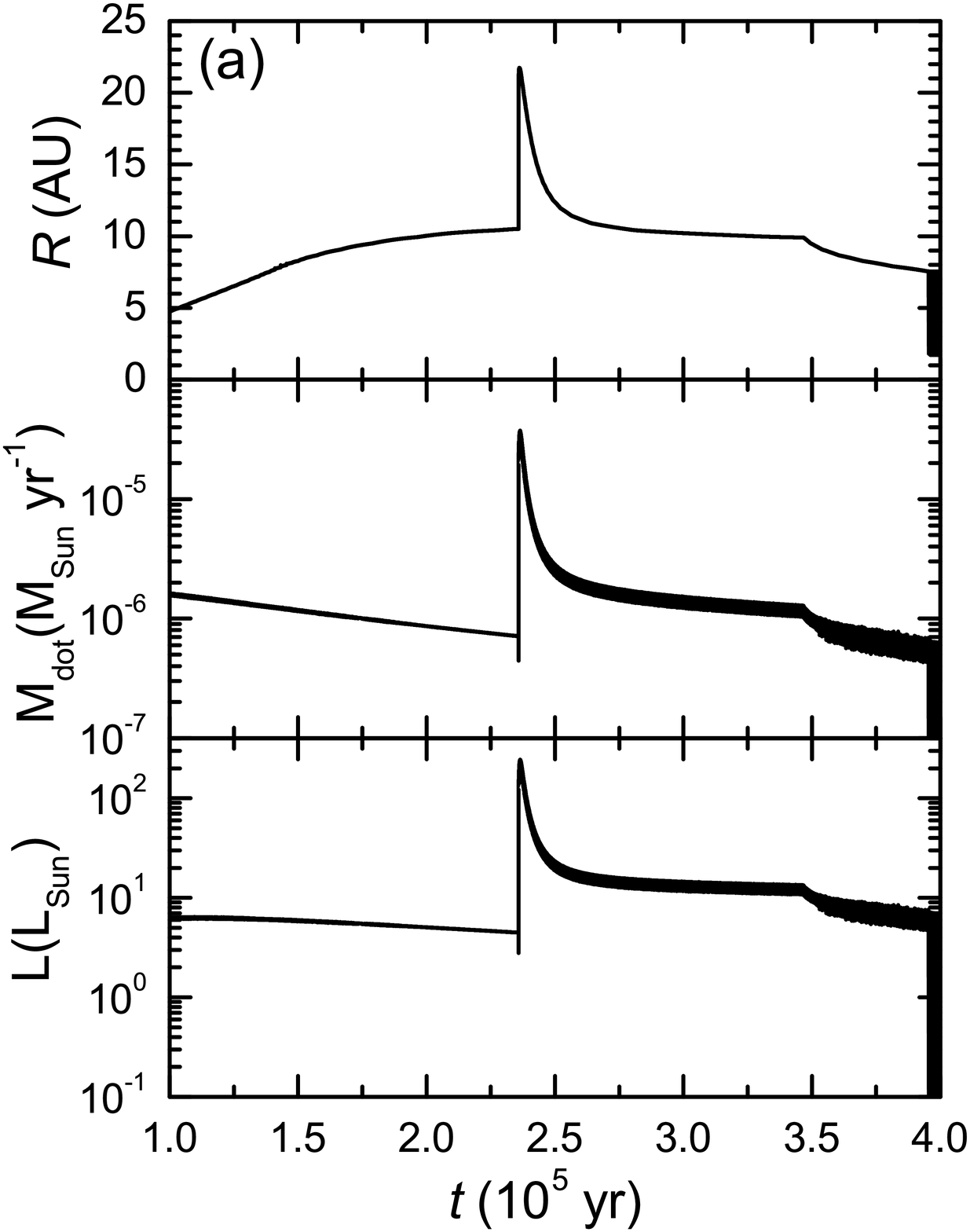}
   \includegraphics[width=4cm]{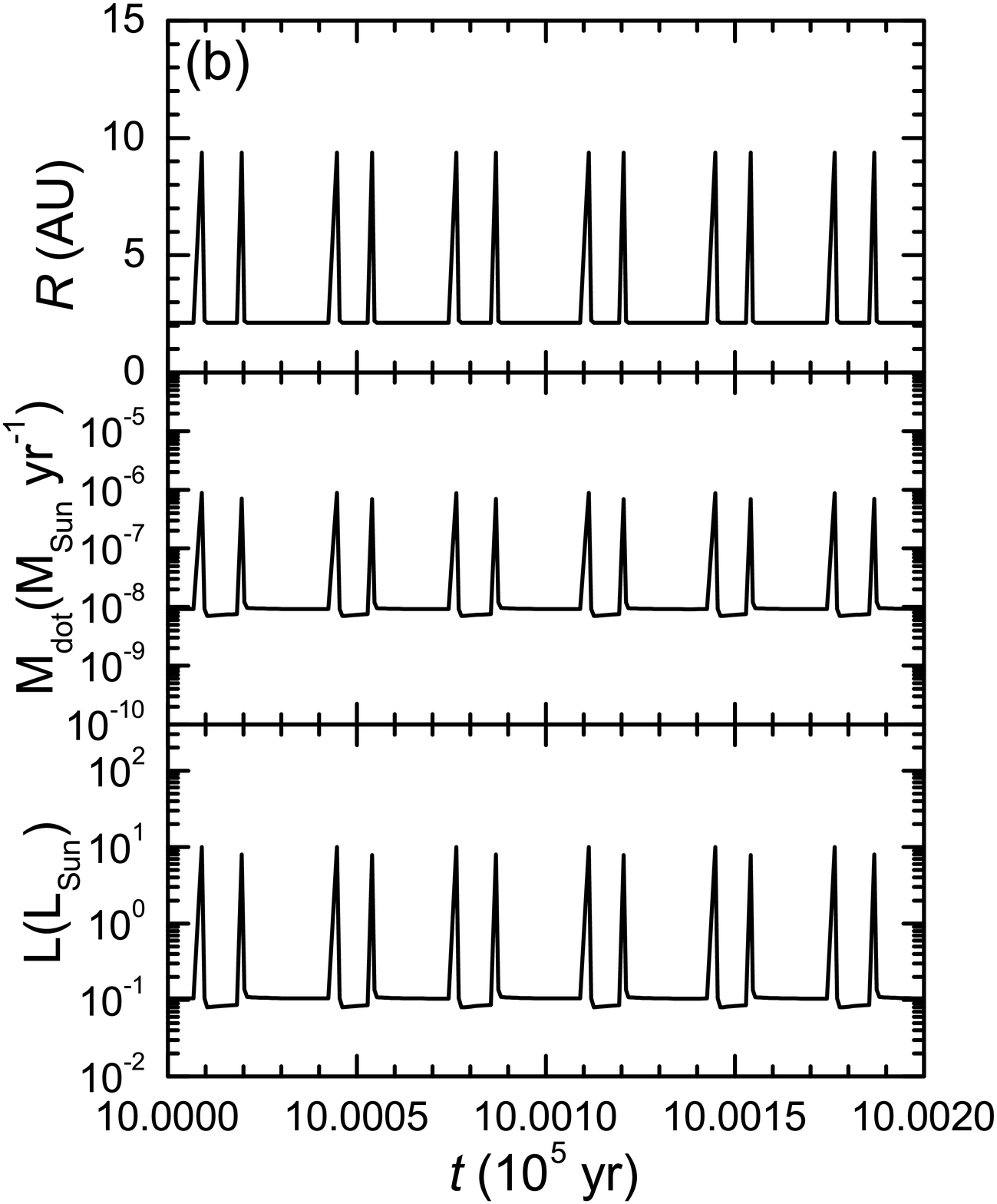}
   \includegraphics[width=4cm]{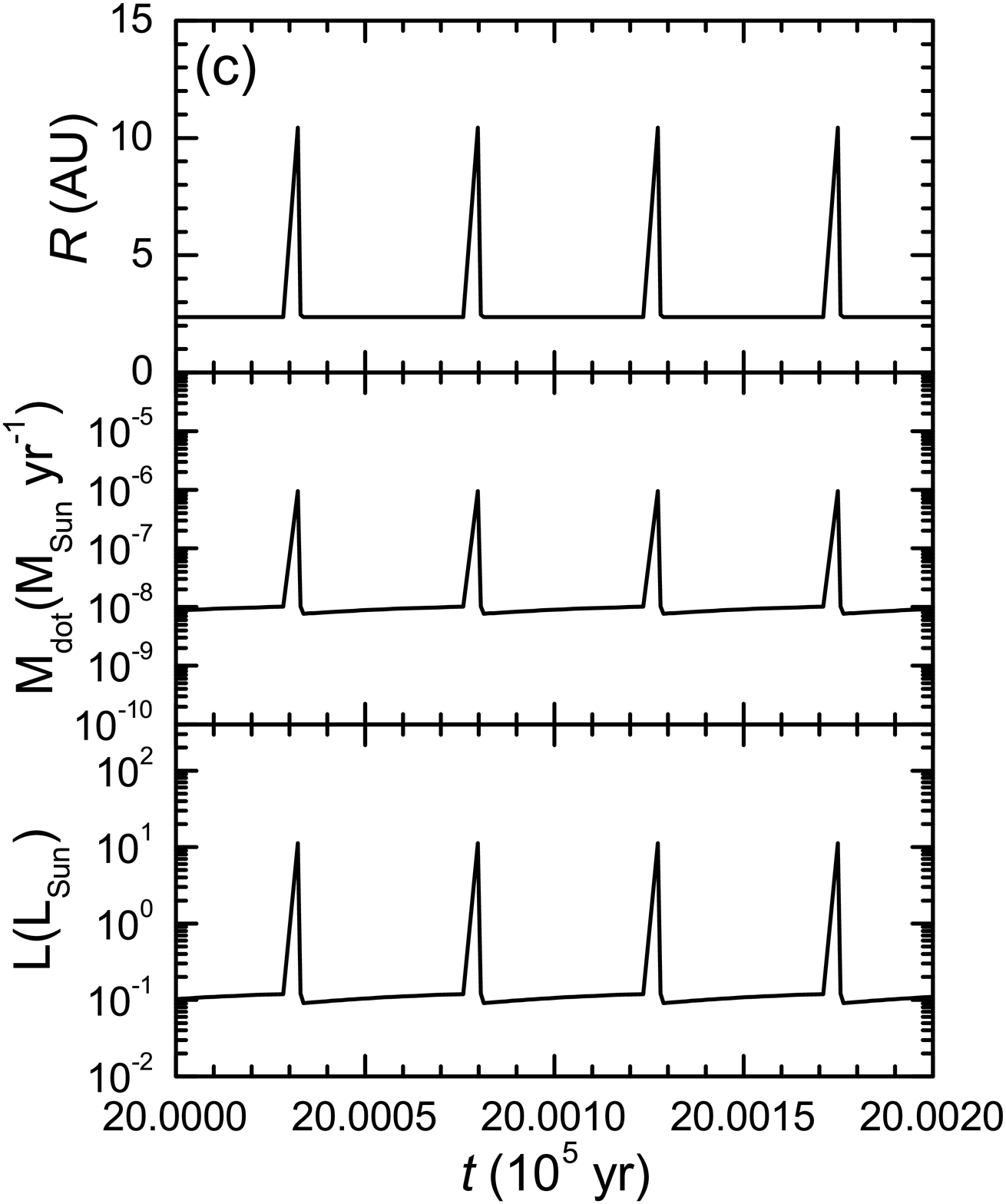}
   \includegraphics[width=4cm]{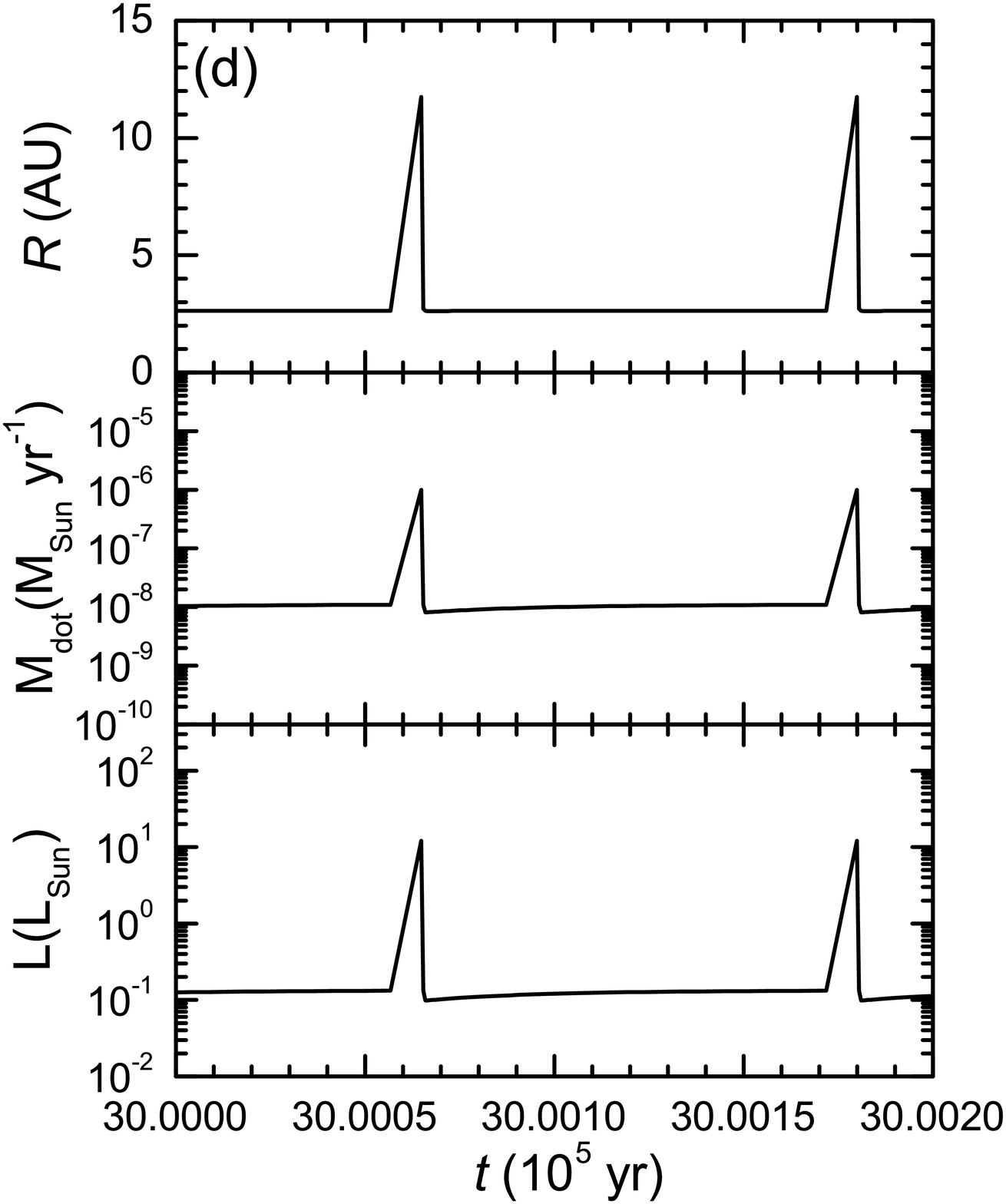}

   \caption{Evolution of the snowline, disk accretion rate, and accretion luminosity for the case with $\omega=3.0 \times 10^{-14}\ \rm s^{-1}$ and $\alpha_{\rm hy}=5\times10^{-3}$. Panel (a) shows the evolution from $t=1.0\times10^5$ to $4.0\times10^5$ yr. The beginning time of panel (b), (c), and (d) are 1.0, 2.0, and 3.0 Myr, respectively.
   }
         \label{fig-mdot}
   \end{figure}

The time durations of the instabilities as functions of $\omega$ for $\alpha_{\rm hy}=5\times 10^{-3}$
and $\alpha_{\rm hy}=5\times 10^{-4}$ are plotted in Fig. \ref{Figdur}.
{We only calculate the time interval when the disk maintains in the unstable stage.
The time of the instability during the episodic phase is not included.}
When $\omega$ increases, the materials in the disks increase and this favors the maintain of the instability.
So the duration of the instability increases.
When $\omega>8-9\ \times 10^{-14}\ \rm s^{-1}$, the duration of the instability decreases slowly.
This is because if the angular velocity is too large, the surface density does not change much at large radius and
the disk tends to be stable.

Compare the cases $\alpha_{\rm hy}=5\times 10^{-3}$ and $\alpha_{\rm hy}=5\times 10^{-4}$, we find that if $\omega<4.5 \times 10^{-14}\ \rm s^{-1}$,
the durations of the instabilities are general the same.
If $\omega$ is larger, the duration is slightly longer for $\alpha_{\rm hy}=5\times 10^{-4}$.
Therefore, for different $\alpha_{\rm hy}$ in our case, both the stabilities and the durations of the instabilities of the disks do not change much.

\section{Influence on the snowline}
When a disk becomes gravitationally unstable,
the viscosity of the disk increases
and the temperature of the disk increases.
The snowline in the disk,
where the temperature at a radius equals $170\ \rm K$,
increases quickly to large radius.
As the material in the disk decreases due to the increase in viscosity,
the disk becomes stable,
the temperature and the radius of the snowline decreases quickly.
So the snowline in the disk reflects the behavior
of the instability.

Fig. \ref{Figsnowline} shows the evolution of the snowline
for different $\omega$ with $\alpha_{\rm hy}=5\times10^{-4}$.
Fig. \ref{Figsnowline}(a), (b), (c), and (d)
show the evolutions for $\omega=1 \times 10^{-14}\ \rm s^{-1}$,
$\omega=3 \times 10^{-14}\ \rm s^{-1}$,
$\omega=6 \times 10^{-14}\ \rm s^{-1}$,
and $\omega=10 \times 10^{-14}\ \rm s^{-1}$, respectively.
The angular velocity in Fig. \ref{Figsnowline}(a) is less than the critical value.
The snowline moves outward due to the influx materials from the cloud core first and achieves a maximum value,
and then moves inward.

In Fig. \ref{Figsnowline}(b),
the snowline increases first as in Fig. \ref{Figsnowline}(a).
When it locates at about 10.5 AU and $t \sim 2.4\times10^5$ years, the disk becomes unstable.
The temperature of the disk increases quickly due to the increase of the viscosities and the accretion luminosity.
The snowline increases quickly to $\sim$ 22 AU.
Then the temperature decreases quickly due to the decrease of surface densities,
so the snowline decreases quickly.
The disk remains unstable during this period.
The snowline begins vibrating quickly at $t=4 \times 10^5\ \rm yr$.
This is because when the disk becomes unstable,
the large viscosity decreases the materials quickly.
As there is not enough mass in the disk,
the disk becomes stable quickly.
The disk is near the critical state of the instability.
As the disk evolves, there is more materials at large radius,
it becomes unstable again.
When $t\sim 3.5$ Myr, the mass of the disk is so low that it remains stable.

The general behaviors of the snowlines in Figs. \ref{Figsnowline} (c) and (d) are similar to that in Fig. \ref{Figsnowline} (b).
The main difference is that as $\omega$ increases, the time duration that the snowline remains at large radius increases and the last of the vibration period increases, which means the time duration of the instability increases and the timescale that a disk is at the critical state of the instability increases.

For all the unstable disks in Fig. \ref{Figsnowline}, the disks become unstable when $t\sim 2 \times 10^5\ \rm yr$,
which is less than the timescale of collapse of the molecular clouds ($\sim 3.3 \times 10^5\ \rm yr$).
For $\omega=1.3 \times 10^{-14}\ \rm s^{-1}$, however, the disk becomes unstable when $t\sim 9 \times 10^5\ \rm yr$,
which is larger than the collapse timescale.
Thus there are cases that the disks becomes unstable at very early stage of disk evolution,
which is consistent with the observations \citep{oso03,rod05,eis05}.

{To show the detail of the snowline, in Fig. \ref{fig-mdot}, we plot the behavior of it at four
time intervals.
We adopt the same parameters as Fig. \ref{Figsnowline} (b). 
We also show the disk accretion rate and the luminosity caused by the mass accretion.
Fig. \ref{fig-mdot} (a) shows the evolution of the snowline and the disk accretion rate from $1.0\times10^5$ to $4\times10^5$ years.
When the disk becomes gravitationally unstable at $t \sim 2.4\times10^5$ years,
the radius of the snowline increases quickly from around 10.5 AU to more than 20 AU.
The disk accretion rate also increases quickly from
$7\times10^{-7}$ to $4\times10^{-5}\ M_{\odot}\ \rm yr^{-1}$.
The corresponding luminosity increases from 4 to 240 $L_\odot$,
where $L_\odot$ is the luminosity of the Sun.
After then, they all decrease with time.
Fig. \ref{fig-mdot} (b) shows the behavior of the snowline beginning at
$t = 1.0\times10^6 $ years.
The location of the snowline is mainly at about 2 AU and increases to about 10 AU for a short period.
The disk accretion rate changes between $\sim 10^{-8}$ and
$\sim 10^{-6}\ M_{\odot}\ \rm yr^{-1}$.
The accretion luminosity changes between 0.1 and 10 $L_\odot$.
The behaviors of the snowline and the disk accretion rate
around $t = 2.0\times10^6 $ (Fig. \ref{fig-mdot} (c)) and $t = 3.0\times10^6 $ (Fig. \ref{fig-mdot} (d)) show similar trends.}

EXors (named after the prototypes EX Lupi) are pre-main sequence stars, which exhibit a brightness increase of a few magnitudes in about several months \citep{her89,aud10,aud14}.
For example, the luminosity of V1118 Orionis increases from 1-2 L$_\sun$ during quiescence to $\sim$7 L$_\sun$ during an outburst.
The corresponding mass accretion rate is estimated to be from about $2.5\times10^{-7}$ to $1.0\times10^{-6} {\rm\ M_\sun\ yr}^{-1}$ \citep{aud10}.
The durations of the eruptions are about 1-2 years, and the outburst usually repeats in a few years \citep{her08}.
Recent observations and simulations found that EXor objects do not display outflows which usually occur in FUor sources \citep{cie18}.
Therefore FUors are believed to experience an early stage of disk evolution, while EXors represent a later stage.
The above results are based on a small sample with 3 (or 4) FUors and 4 EXors.
More observations and theoretical works will give a clearer view in the future.
Fig. \ref{fig-mdot} (b-d) show similar amplitudes and magnitudes of luminosity as EXors.
The accretion luminosity increases from $\sim 10^{-1}$ L$_{\sun}$ in quiescence to $\sim$ 10 L$_{\sun}$ for eruptions.
But the periods between two outbursts are from $\sim 10$ to $\sim$100 years, which are longer than the duty cycles of EXors.
Subsequent research with a detailed model will address this issue in the future.

\section{discussion and Conclusions}
In this paper, we calculate the evolutions of the disks form from the collapse of the molecular cloud cores.
We fix the temperature and the mass of the cloud cores and change the angular velocities smoothly to
investigate the dependence of the gravitational stabilities of the disks on the angular velocities of the cloud cores.
We also study how the hydrodynamic viscosity in the dead zone affects the instability of the disks.

{The differences of our work with Vorobyov and Basu's are as follows.
First, we adopt a collapse model of molecular cloud cores,
while they used specific initial surface densities of disks to stand for the cloud cores.
Second, we use a more realistic viscosity structure in the disk and investigate the effect of hydrodynamic viscosities on the disk instability,
while they use a constant $\alpha$ value in all their papers.
We also have a lower inner radius (0.3 AU) than their model (6 AU),
which will affect the global evolution of the disk.}

We first fix the hydrodynamic viscosity to be $\alpha_{\rm hy}=5\times10^{-4}$.
We find that when the angular velocities increase,
the disks tend to become unstable.
For molecular cloud cores with $T=15\ \rm K$ and $M=1\ M_{\odot}$,
the disks become unstable when the angular velocity, $\omega$,
 is larger than $1.2 \times 10^{-14}\ \rm s^{-1}$.
If $\omega$ increases, the time duration of the instability increases if $\omega<\sim7-8 \times 10^{-14}\ \rm s^{-1}$ and the timescale that the disks are at the critical state of the instability increases.

We also change the hydrodynamical viscosity from $\alpha_{\rm hy}=5\times10^{-4}$ to $\alpha_{\rm hy}=5\times10^{-3}$ of the disk.
The change of the viscosity does not affect the trigger of the instability much.
The time durations of the instabilities for both $\alpha_{\rm hy}$ do not change much either.
The main effects are on the timescale of the disks that are at the critical state of the instability.

{When $\alpha_{\rm hy}=5\times10^{-4}$, for $\omega>\sim3 \times 10^{-14}\ \rm s^{-1}$,
the disks become unstable at $t \sim 2\times10^5$,
which is less than the collapse timescale of molecular cloud core and consistent with the observations
\citep{oso03,rod05,eis05}. When $\alpha_{\rm hy}=5\times10^{-3}$, the disks become unstable at later time.}


\acknowledgments
The support provided by China Scholarship Council (CSC) during a visit of Ning Sui (No.201706175038) to MPIA is acknowledged. This work is supported in part by the National Natural Science Foundation of China (no. 11273013). Min Li is supported by the College of Sciences at the University of Nevada and NASA grant NNX16AK08G leaded by  Jason H. Steffen. Finally, we thank the referee for comments that clarified aspects of the paper.


\clearpage
\end{document}